\pdfoutput=1

\RequirePackage{ifpdf}
\documentclass[12pt,nohyper]{JHEP3}

\usepackage{epsfig}
\usepackage{float}
\usepackage{amsmath}
\usepackage{array}
\usepackage{cite}
\usepackage{arydshln}

\let\normalcolor\relax

\newcommand{\smallplus}{\hspace{0.5pt}\text{{\small+}}\hspace{-0.5pt}}

\newcommand{\pl}{\smallplus}
\newcommand{\la}{\langle}
\newcommand{\ra}{\rangle}

\title{\hspace{-0.0cm}{\LARGE Into the Amplituhedron}}
\author{\vspace{-.5cm}Nima Arkani-Hamed$^{a}$ and Jaroslav Trnka$^{b}$\\
{\footnotesize{\it $^{a}$ School of Natural Sciences, Institute for Advanced Study, Princeton, NJ 08540, USA}\\
{\it $^{b}$ California Institute of Technology, Pasadena, CA 91125,
USA}}\vspace{-.5cm}} \preprint{2013}

\abstract{We initiate an exploration of  the physics and geometry of
the amplituhedron, starting with  the simplest case of the integrand
for four-particle scattering in planar ${\cal N} = 4$ SYM.  We show
how the textbook structure of the unitarity double-cut  follows from
the positive geometry. We also use the geometry to expose the
behavior of the multicollinear limit,  providing a direct motivation
for studying the logarithm of the amplitude. In addition to
computing the two and three-loop integrands, we explore various
lower-dimensional faces of the amplituhedron, thereby computing
non-trivial cuts of the integrand to all loop orders. }

\preprint{CALT-68-2873}

\begin{document}

\newpage
\section{Geometry and Physics of the Amplituhedron}

In \cite{P1}, we introduced a new geometric object--the
Amplituhedron--underlying the physics of scattering amplitudes for
${\cal N}=4$ SYM in the planar limit. At tree level, the
amplituhedron is a natural generalization of
``the inside of a convex polygon".  Loops arise by extending the
geometry to incorporate the idea of ``hiding particles" in the only
natural way possible.

The amplituhedron ${\cal A}_{n,k,L}$ for $n$-particle N$^k$MHV
amplitudes at $L$ loops, lives in $G(k,k+4;L)$, which is the space
of $k$-planes $Y$ in $k+4$ dimensions, together with $L$ 2-planes
${\cal L}_1, \cdots, {\cal L}_L$  in the $4$ dimensional complement
of $Y$. The external data are given by a collection of $n$ $(k+4)$
dimensional vectors $Z_a^I$. Here $a=1, \cdots n$, and $I = 1,
\cdots, (k+4)$. This data is taken to be ``positive", in the sense
that all the ordered $(k+4) \times (k+4)$ determinants $\langle
Z_{a_1} \cdots Z_{a_{k+4}} \rangle > 0$ for $a_1 < \cdots <
a_{k+4}$. The subspace of ${\cal A}_{n,k,L}$ of $G(k,k+4;L)$ is
determined by a ``positive" linear combination of the (positive)
external data.  The $k$-plane is $Y_{\alpha}^I$, and the 2-planes
are ${\cal L}_{\gamma (i)}^I$, where $\gamma = 1,2$ and  $i = 1,
\dots, L$ . The amplituhedron is the space of all $Y, {\cal
L}_{(i)}$ of the form
\begin{equation}
Y_{\alpha}^I = C_{\alpha a} Z_a^I, \qquad {\cal L}_{\gamma (i)}^{I}
= D_{\gamma a (i)} Z_a^I
\end{equation}
where the $C_{\alpha a}$ specifies a $k$-plane in $n$-dimensions,
and the $D_{\gamma a (i)}$ are $L$ 2-planes living in the $(n-k)$
dimensional complement of $C$, with the positivity property that for
any $0 \leq l \leq L$, all the ordered maximal minors of the $(k + 2
l) \times n$ matrix
\begin{equation}
\left(\begin{array}{ccc} & D_{(i_1)} & \\
\hdashline & \vdots &  \\ \hdashline & D_{(i_l)} & \\ \hdashline & C
\end{array} \right)
\end{equation}
are positive.

There is a canonical rational form $\Omega_{n,k;L}$
associated with ${\cal A}_{n,k;L}$, with the property of having
logarithmic singularities on all  the lower-dimensional boundaries
of ${\cal A}_{n,k;L}$. The loop integrand form for the
super-amplitude is naturally extracted from $\Omega_{n,k;L}$
\cite{P1}.

The amplituhedron can be defined in a few lines, as we have just
done. But the resulting geometry is incredibly rich and
intricate--as it must be, to generate all the structure  found in
planar ${\cal N} = 4$ SYM scattering amplitudes to all loop orders!
For instance, the singularity structure of the amplitude is
reflected in the geometry of the various boundaries of the
amplituhedron; studying this geometry in some of the simplest cases
allows us to see the emergence of locality and unitarity from
positive geometry.

Even just the tree amplituhedron generalizes the positive
Grassmannian $G_+(k,n)$ \cite{alex}. A complete understanding of $G_+(k,n)$
revealed many surprising connections to other structures, from the
fundamentally combinatorial backbone of affine permutations, to
cluster algebras, to the physical connection with on-shell processes
\cite{alex, FG, positive}.  It is natural to expect the full amplituhedron
${\cal A}_{n,k,L}$ to have a much richer structure. A complete
understanding of the full geometry of the amplituhedron, at the same
level as our understanding of the positive Grassmannian, will likely
involve further physical and mathematical ideas. Our goal in this
note is to begin laying the groundwork for this exploration, by
looking at various simple aspects of amplituhedron geometry in the
simplest non-trivial case of clear physical interest.

While the tree amplituhedron generalizes the positive Grassmannian
in a direct way, extending the notion of positivity to external
data, the extension of positivity associated with ``hiding
particles" which gives rise to loops is more novel and interesting.
The very simplest case of four-particle scattering has $k=0, n=4$.
Here, we don't have the additional structure of Grassmann components
for the external data \cite{P1}, the external data are just the
ordinary bosonic momentum-twistor\cite{A1} variables
$Z^I_1,Z^I_2,Z^I_3,Z^I_4$, for $I=1, \cdots, 4$. Furthermore, the
constraint of positivity for external data is trivial in this case;
indeed using a $GL(4)$ transformation we can set the $4 \times 4$
matrix $(Z_1, \cdots, Z_4)$ to identity.  The loop variables are
just lines in momentum-twistor space (or better, two-planes in
four-dimensions), which correspond to points in the (dual)
space-time. Having set the $Z$ matrix to the identity, each $2
\times 4$ matrix for the lines ${\cal L}^I_{\gamma (a)}$ is simply
identified with the $D$ matrices $D_{(i)}$.

The amplituhedron positivity
constraints are that the all the ordered minors of each $D_{(i)}$
matrix are positive
\begin{equation}
(12)_i, (13)_i, (14)_i, (23)_i, (24)_i,(34)_i > 0
\end{equation}
We also have mutual positivity, that the $4 \times 4$ determinant
$\langle D_{(i)} D_{(j)} \rangle > 0$, which tells us that
\begin{align}
(12)_i (34)_j + (23)_i (14)_j + (34)_i
(12)_j + (14)_i (23)_j - (13)_i (24)_j - (24)_i (13)_j > 0
\end{align}
We can also express these conditions in a convenient gauge, where
\begin{equation}
D_{(i)} = \left(\begin{array}{cccc} 1 & x_i & 0 & -w_i \\ 0 &
y_i & 1 & z_i
\end{array} \right)
\end{equation}
Then the positivity of each $D_{(i)}$ simply tells us that
\begin{equation}
x_i,y_i,z_i,w_i > 0
\end{equation}
while the mutual positivity conditions become
\begin{equation}
(x_i - x_j)(z_i - z_j) + (y_i - y_j)(w_i - w_j)<0
\end{equation}

In this note we study various aspects of  the geometry defined by these
inequalities, as well as the corresponding canonical form $\Omega$,
which directly gives us the loop integrand for four-particle scattering. Of course the
four-particle amplitude has been an object of intensive study
for many years \cite{Bern:2006ew, Bern:2005iz, Bern:2007ct,
Bourjaily:2011hi, Eden:2012tu}, with loop integrand now available
through seven loops. But our approach will be fundamentally
different from previous works. We will not begin by drawing planar
diagrams made out of ``boxes", we will make no mention of recursion
relations, we will not make ansatze for the integrand which are
checked against cuts, and we will make no mention of physical
constraints from exponentiation of infrared divergences etc.
Instead, we will discover all the known general properties of the
loop integrand, and many other properties besides, directly by
studying the positive geometry of the amplituhedron.

We will start with a lightning review of the one-loop geometry,
which is just that of $G_+(2,4)$, mostly to define some notation and
nomenclature. We then do some warm-up exercises for associating
canonical forms $\Omega$ with spaces specified by particularly
simple inequalities, which will come in handy in later sections. The
first non-trivial case with mutual positivity is obviously two
loops, and we show how to triangulate the space and extract the
loop integrand, matching the well-known result given as a
sum of two double-boxes. Interestingly, while our triangulation of
the two-loop amplituhedron is manifestly ``positive", the sum of
double-boxes is not, with each term having singularities outside the amplituhedron
that only cancel in the sum.

We then make some general observation on the structure of certain
cuts of the amplitude, which correspond to various boundaries of the
amplituhedron. In particular, the textbook understanding of
unitarity as following from the break-up of the loop integrand into two
parts sewed together on the ``unitarity cut" follows in a beautiful way from
positive geometry. These general
results and some further explicit triangulations also allow us to
determine the three-loop integrand. We move on to exploring another
natural set of cuts that take the amplitude into the multi-collinear
region. This exposes a fascinating property of cuts of the
multi-loop integrand: the residues depend not only on
the final cut geometry, but also on the path taken to reach that
geometry. Studying the combinatorics of this path dependence naturally
motivates looking at the logarithm
of the amplitude, and explains why the log has such good IR
behavior.

From our new perspective, the determination of the integrand to all loop orders
requires a complete understanding of the full
amplituhedron geometry. We have not yet achieved this yet, but we
believe that a systematic approach to this problem is
possible. As a prelude,  we give a survey of
some of the lower-dimensional ``faces" of amplituhedron. We can
explicitly triangulate these faces and find their corresponding
canonical forms, which give us cuts of the full integrand. This
already gives us highly non-trivial all-loop order information about
the integrand, in many cases not readily available from any other approach.

\section{One Loop Geometry}

At one loop we have a single line ${\cal L}_1 {\cal L}_2$, which we
often also called ``$(AB)$".  The geometry is given by the positive
Grassmannian $G_+(2,4)$. The external data form a polygon in
$\mathbb{P}^3$ with vertices $Z_1$, $Z_2$, $Z_3$, $Z_4$ and edges
$Z_1Z_2$, $Z_2Z_3$, $Z_3Z_4$, $Z_1Z_4$.
$$
\includegraphics[scale=.65]{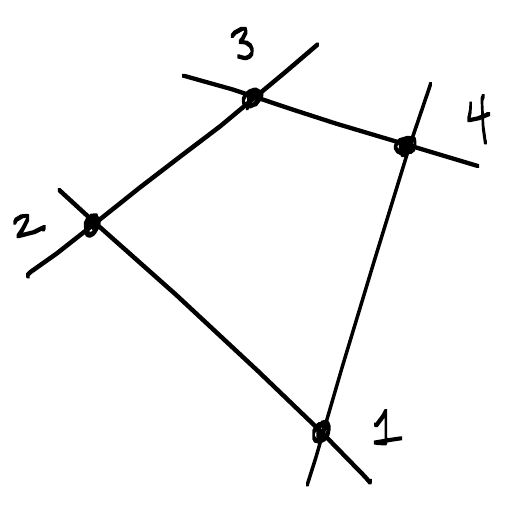}
$$
The line $AB={\cal L}_1 {\cal L}_2$ is parametrized as
\begin{equation}
{\cal L}^I_\gamma = D_{\gamma a}Z_a^I
\end{equation}
where $\gamma=1,2$ and $a,I=1,\dots,4$. The matrix $D$ represents a
cell of positive Grassmannian $G_+(2,4)$, in the generic case it is a
top cell. In one particularly convenient gauge-fixing we can write
\begin{equation}
D = \left(
      \begin{array}{cccc}
        1 & x &0 & -w \\
        0 & y & 1 & z \\
      \end{array}
    \right)\label{Cmatrix}
\end{equation}
where $x,y,z,w>0$. This gauge-fixing of the $D$ matrix covers all
boundaries by sending variables $x,y,z,w$ to zero or infinity.

The form with logarithmic singularities on the boundaries of the
space is trivially
\begin{equation}
\Omega = \frac{dx}{x}\frac{dy}{y}\frac{dw}{w}\frac{dz}{z}
\end{equation}
The  boundaries occur when one of the variables approaches $0$ or
$\infty$. We can easily translate this expression back to
momentum twistor space by solving two linear equations:
\begin{equation}
Z_A = Z_1 + xZ_2 - w Z_4,\qquad Z_B = yZ_2 + Z_3 + zZ_4
\end{equation}
which gives
\begin{equation}
\Omega = \frac{\la AB\,d^2Z_A\ra\la AB\,d^2Z_B\ra\la1234\ra^2} {\la
AB12\ra\la AB23\ra\la AB34\ra\la AB14\ra}
\end{equation}

We now describe the boundaries of this space in detail--these are
nothing but all the cells of $G_+(2,4)$, which have also been
described at length in e.g. \cite{positive}. We describe them in
detail here since the same geometry will arise repeatedly in the context of cuts of
the multiloop amplitudes. At the level of the form they correspond
to logarithmic singularities. In giving co-ordinates for the boundaries,
we will freely use different gauge-fixings as convenient for
any given case, with all parameters positive. They will  always be
trivially related to boundaries of (\ref{Cmatrix}).

The first boundaries occur when line
$AB$ intersects one of the lines $Z_1Z_2$, $Z_2Z_3$, $Z_3Z_4$ or
$Z_1Z_4$. In the gauge-fixing (\ref{Cmatrix}) this sets one of the
variables to $x,y,z,w$ to $0$. In particular,
\begin{equation}
\la AB12\ra =
w,\quad \la AB23\ra = z,\quad \la AB34\ra = y,\quad \la AB14\ra = x
\end{equation}
where we suppressed $\la1234\ra$. For cutting $Z_1Z_2$, $\la AB12\ra
= w=0$ we get
$$
\begin{minipage}[c]{0.23\textwidth}
\includegraphics[scale=.65]{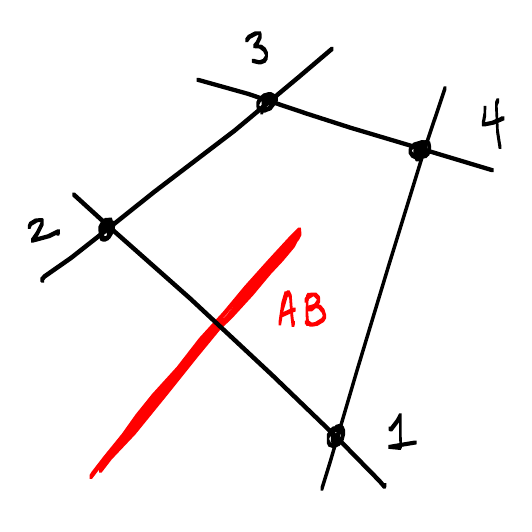}
\end{minipage}
\begin{minipage}[c]{0.2\textwidth}
$$\left(
      \begin{array}{cccc}
        1 & x & 0 & 0 \\
        0 & y & 1 & z \\
      \end{array}
    \right)$$
\end{minipage}
$$
In all four cases the form is the dlog of remaining three variables;
we will suppress writing it explicitly.

The second boundaries occur when the line $AB$
intersects two lines $Z_iZ_{i\pl1}$ and $Z_j Z_{j \pl 1}$. There are two
distinct cases. If we cut two non-adjacent lines $Z_1Z_2$, $Z_3Z_4$
or $Z_2Z_3$, $Z_1Z_4$ there is just one solution. For the first one
$\la AB12\ra=\la AB34\ra=0$ we have
$$
\begin{minipage}[c]{0.23\textwidth}
\includegraphics[scale=.65]{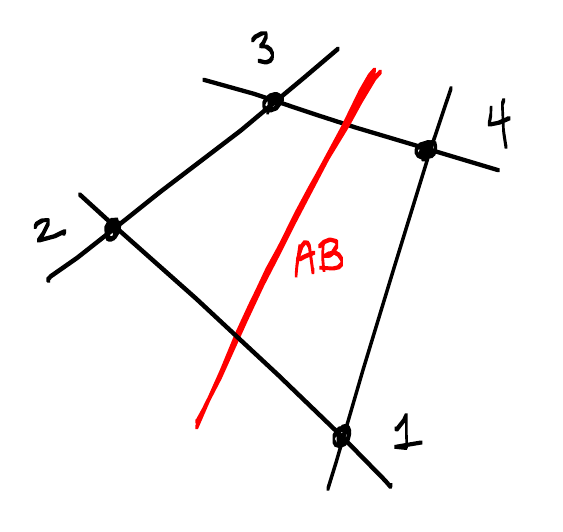}
\end{minipage}
\begin{minipage}[c]{0.2\textwidth}
$$\left(
      \begin{array}{cccc}
        1 & x & 0 & 0 \\
        0 & 0 & 1 & z \\
      \end{array}
    \right)\qquad
$$
\end{minipage}
$$
In the second case we intersect two adjacent lines. Let us cut
$Z_1Z_2$, $Z_2Z_3$ (the other three cases are cyclically related), ie.
$\la AB12\ra=\la AB23\ra=0$. There are two different solutions --
either the line $AB$ passes through $Z_2$ or the line $AB$ lies in
the plane $(Z_1Z_2Z_3)$.
$$
\begin{minipage}[c]{0.23\textwidth}
\includegraphics[scale=.65]{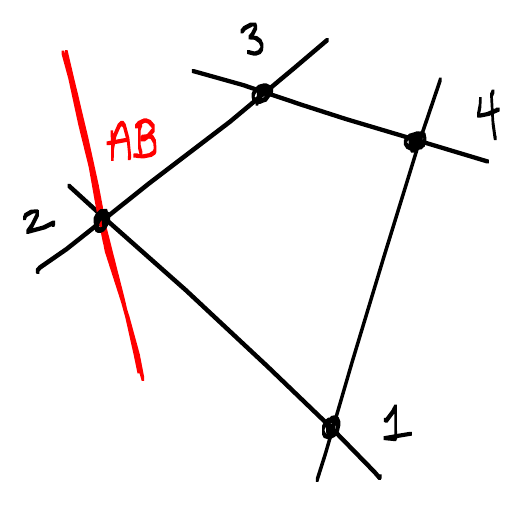}
\end{minipage}
\begin{minipage}[c]{0.2\textwidth}
$$\left(
      \begin{array}{cccc}
        1 & 0 & 0 & 0 \\
        0 & y & 1 & z \\
      \end{array}
    \right)
$$
\end{minipage}\qquad
\begin{minipage}[c]{0.23\textwidth}
\includegraphics[scale=.65]{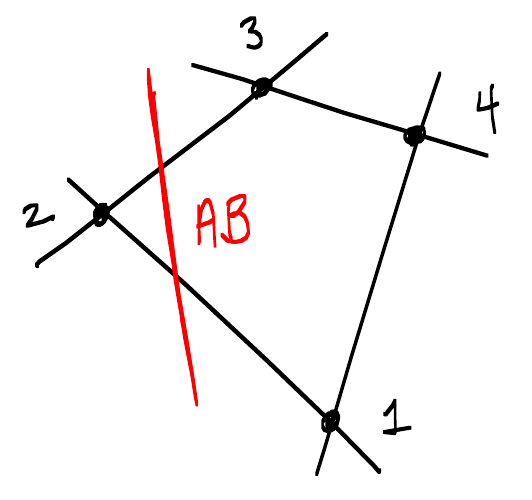}
\end{minipage}
\begin{minipage}[c]{0.2\textwidth}
$$
\left(
      \begin{array}{cccc}
        1 & x & 0 & 0 \\
        0 & y & 1 & 0 \\
      \end{array}
    \right)
$$
\end{minipage}
$$
There are two different types of third boundaries. The first type is
a triple cut -- the line $AB$ intersects three of four lines. One
representative is $\la AB12\ra = \la AB23\ra= \la AB34\ra = 0$.
There are two solutions to this problem. Either $AB$ passes through
$Z_2$ and intersects the line $Z_3Z_4$ or $AB$ passes through $Z_3$
and intersects the line $Z_1Z_2$.
$$
\begin{minipage}[c]{0.23\textwidth}
\includegraphics[scale=.65]{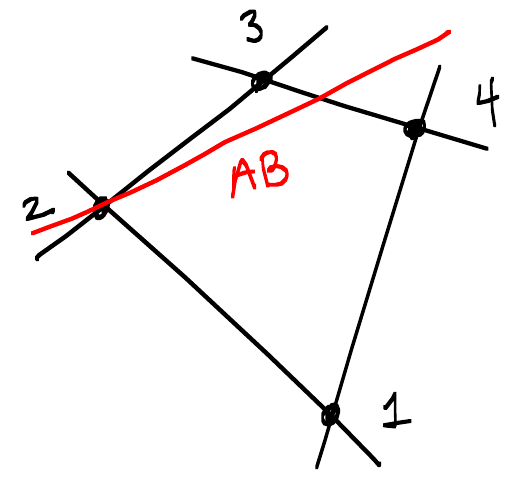}
\end{minipage}
\begin{minipage}[c]{0.2\textwidth}
$$\left(
      \begin{array}{cccc}
        0 & 1 & 0 & 0 \\
        0 & 0 & 1 & \alpha \\
      \end{array}
    \right)
$$
\end{minipage}\qquad
\begin{minipage}[c]{0.23\textwidth}
\includegraphics[scale=.65]{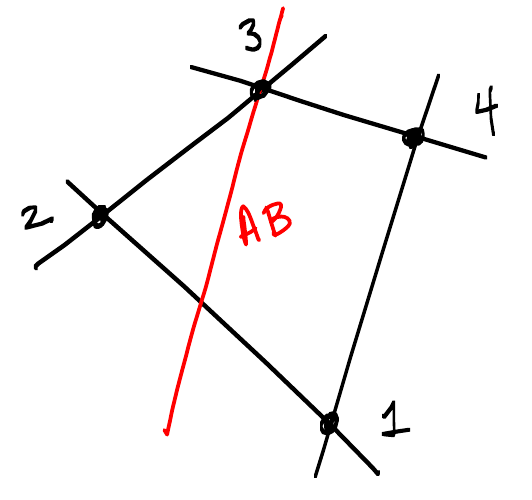}
\end{minipage}
\begin{minipage}[c]{0.2\textwidth}
$$\left(
      \begin{array}{cccc}
        0 & 0 & 1 & 0 \\
        -\alpha & -1 & 0 & 0 \\
      \end{array}
    \right)
$$
\end{minipage}
$$
There is also a ``composite" cut when we cut only two lines while
imposing three constraints. We can pass $AB$ through $Z_2$  while lying in the plane
plane $(Z_1Z_2Z_3)$.
$$
\begin{minipage}[c]{0.23\textwidth}
\includegraphics[scale=.65]{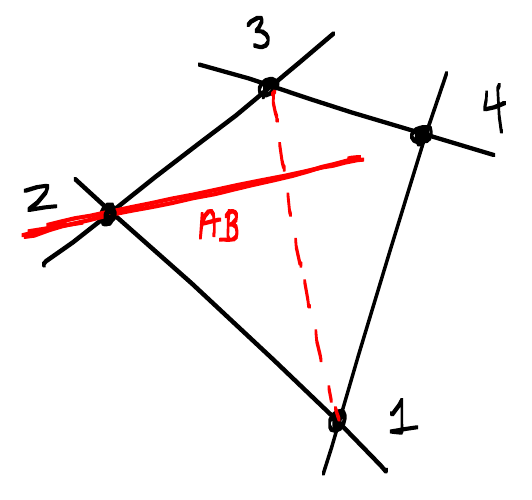}
\end{minipage}
\begin{minipage}[c]{0.2\textwidth}
$$\left(
      \begin{array}{cccc}
        0 & 1 & 0 & 0 \\
        -\alpha & 0 & 1 & 0 \\
      \end{array}
    \right)
$$
\end{minipage}
$$
Finally, for the quadruple cuts we can either cut all four lines
which localizes $AB$ to $AB=Z_1Z_3$ or $AB=Z_2Z_4$, or we can
consider the "composite'' cut $AB=Z_1Z_2$ (and cyclically related)
which cuts only three lines (not $Z_3Z_4$) while still imposing four
constraints.
$$
\begin{minipage}[c]{0.23\textwidth}
\includegraphics[scale=.65]{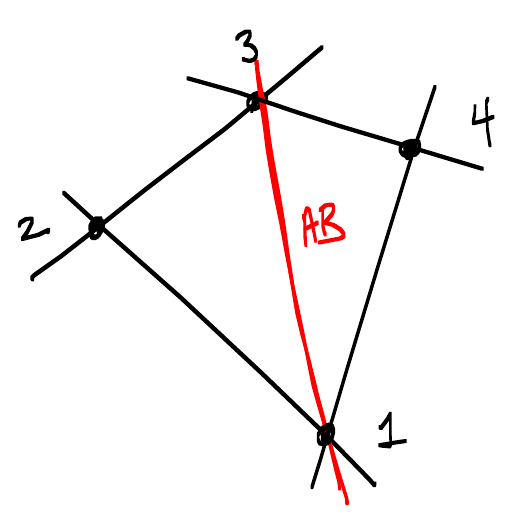}
\end{minipage}
\begin{minipage}[c]{0.2\textwidth}
$$\left(
      \begin{array}{cccc}
        1 & 0 & 0 & 0 \\
        0 & 0 & 1 & 0 \\
      \end{array}
    \right)
$$
\end{minipage}\qquad
\begin{minipage}[c]{0.23\textwidth}
\includegraphics[scale=.65]{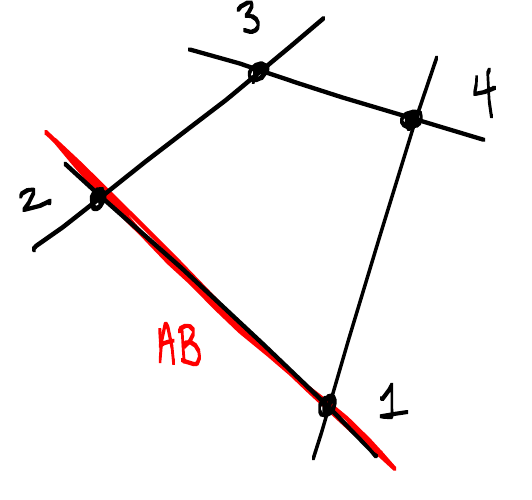}
\end{minipage}
\begin{minipage}[c]{0.2\textwidth}
$$\left(
      \begin{array}{cccc}
        1 & 0 & 0 & 0 \\
        0 & 1 & 0 & 0 \\
      \end{array}
    \right)
$$
\end{minipage}
$$

\section{Warmup Exercises}

The amplituhedron is defined by various positivity conditions. We
will shortly be ``triangulating" the spaces defined by these inequalities and finding the
canonical form $\Omega$ associated with them. But it will be helpful
to practice on some simpler cases, which will also later be useful to
determining amplitudes and cuts of amplitudes.

Let us start with a trivial example; suppose we have
\begin{equation}
a < x < b
\end{equation}
It's obvious that the form is $\frac{1}{x - a} - \frac{1}{x - b}$,
but lets reproduce this in a heavy-handed way, from the viewpoint
using ``positive co-ordinates". In this case, we can write
\begin{equation}
x = a + (b - a)  \frac{\alpha}{1 + \alpha}
\end{equation}
Note that for $\infty > \alpha > 0$, we cover the entire range of $a < x < b$. The canonical form is just
$\frac{d \alpha}{\alpha}$, which can be re-written in the original co-ordinates as
\begin{equation}
\frac{d \alpha}{\alpha} = \frac{dx}{x -a} - \frac{dx}{x - b} =
\frac{(a - b) dx}{(x - a)(x - b)}
\end{equation}

Next, consider $0 < x_1 < x_2$. Once again, we can use positive
variables
\begin{equation}
x_1 = \alpha_1, \qquad x_2 = \alpha_1 +
\alpha_2
\end{equation}
and the form is quite trivially
\begin{equation}
\frac{d \alpha_1}{\alpha_1} \frac{d \alpha_2}{\alpha_2} = \frac{dx_1
dx_2}{x_1 (x_2 - x_1)}
\end{equation}
We will henceforth skip the step of parametrization with
positive variables, and also omit the measure factor in
presenting results.

Next consider $0 < x_1 < x_2 < a$, the form
is
\begin{equation}
\left(\frac{1}{x_1} - \frac{1}{x_1 - a}\right)\left(\frac{1}{x_2 -
x_1} - \frac{1}{x_2 - a}\right) = \frac{a}{x_1 (x_2 - x_1) (a -
x_1)}
\end{equation}
This extends trivially to e.g. $0 < x_1 < x_2 < a < x_3 < x_4 < b$,
for which the form is
\begin{equation}
\frac{a b}{x_1 (x_2 - x_1) (a - x_2) (x_3 - a) (x_4 - x_3)(b - x_4)}
\end{equation}
We will find it convenient to use a notation to represent these
forms. Consider a chain of inequalities of the form $0 < X_1
< X_2 \cdots < X_N$. Some of the $X$'s are the variables our form
depends on, and some are constants like $a,b$ in our previous
examples. We will represent the constants by underlining the
corresponding $X$'s. In this notation, the form accompanying our two
examples above are denoted as $[x_1, x_2, \underline{a}]$ and
$[x_1,x_2,\underline{a},x_3,x_4,\underline{b}]$. As yet another
example,

\begin{align} [x_1, \underline{a},
\underline{b},x_2,x_3,\underline{c},x_4] = \left(\frac{1}{x_1} -
\frac{1}{x_1 - a}\right)\left(\frac{1}{x_2 - b} - \frac{1}{x_2 -
c}\right)\left(\frac{1}{x_3 - x_2} - \frac{1}{x_3 - c}\right)\left(
\frac{1}{x_4 - c}\right)
\end{align}

Next, suppose we have $x_i, y$ with $y > x_i$
for all $i$. Then, if the $x's$ are ordered so that $x_1 < \cdots, <
x_n$, we have $y> x_n$, and the form is
\begin{equation}
[x_1, \cdots, x_n, y] = \frac{1}{x_1} \frac{1}{x_2 - x_1} \cdots
\frac{1}{x_n - x_{n-1}} \frac{1}{y - x_n}
\end{equation}
Then we simply sum over all the permutations
\begin{equation}
\sum_\sigma [x_{\sigma_1}, \cdots, x_{\sigma_n}, y]
\end{equation}
Note that individual terms in this sum have spurious poles $(x_i -
x_j)$, which cancel in the sum. Indeed, in this simple case, it is
trivial to do the sum explicitly, and find
\begin{equation}
\frac{y^{n-1}}{(y - x_1)(y-x_2) \cdots (y - x_n) x_1 \cdots x_n}
\end{equation}
Extremely naively, we may have expected the product in the
denominator, but why is there is a numerator factor? The reason is
that otherwise, the form would not have only logarithmic
singularities! For instance, the residues on $x_1, \cdots, x_n \to
0$ would give $1/y^n$; it is the numerator that makes this $1/y$.
We can extend this to $y_I > x_i$ for a collection of $m$ $y$'s. This
means that the smallest $y$ is larger than the largest $x$. Thus the
form is
\begin{equation}
\sum_{\sigma, p}  [x_{\sigma_1}, \cdots, x_{\sigma_n},y_{p_1}
\cdots, y_{p_m}]
\end{equation}
Again the spurious poles cancel in the sum, but the forms are more
interesting. In the simplest new case where $n=3, m=2$ the form is
\begin{equation}
\frac{x_1 x_2 x_3 y_1 + x_1 x_2 x_3 y_2 - x_1 x_2 y_1 y_2 - x_1 x_3
y_1 y_2 - x_2 x_3 y_1 y_2 + y_1^2 y_2^2}{x_1 x_2 x_3 (y_1 - x_1)(y_1
- x_2)(y_1 - x_3)(y_2 - x_1)(y_2 - x_2)(y_3 - x_3)}
\end{equation}

Let us now consider the inequality $x, y > 0$ and
also $x+ y < 1$, or $x + y > 1$. The first case is just the inside
of a triangle, while the second case is a quadrilateral:
$$
\includegraphics[scale=.75]{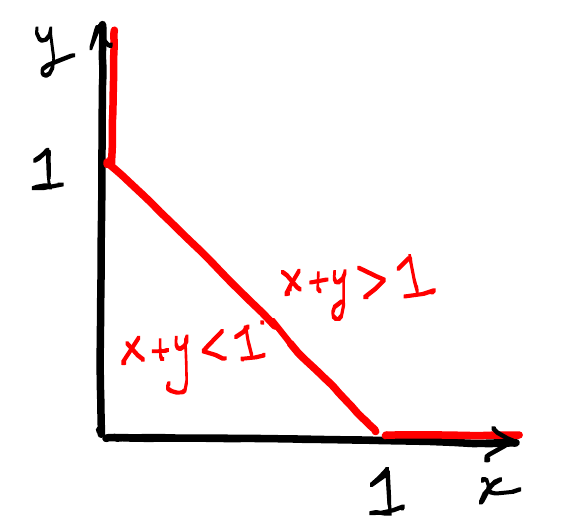}
$$
Obviously the form in the first case $x + y < 1$ is
\begin{equation}
\frac{-1}{x y (x + y - 1)}
\end{equation}
For the second case, the region can be broken into two pieces in
obvious ways. For instance, if $x > a$, there is no further
restriction on $y$, while if $x<1$, we must have $y > 1 - x$
$$
\includegraphics[scale=.75]{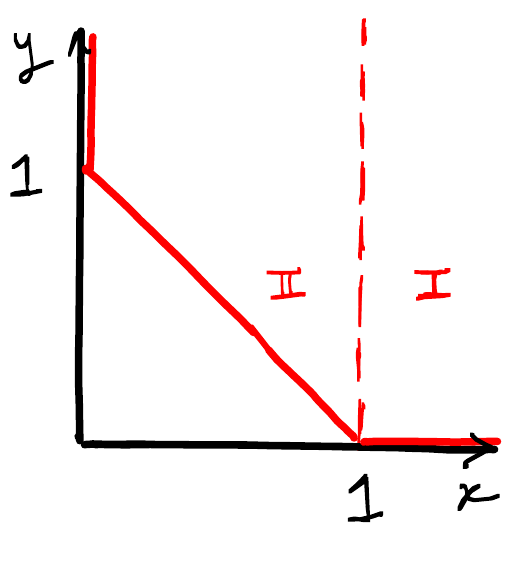}
$$
The form is then
\begin{equation}
\frac{1}{x - 1} \frac{1}{y} + \frac{1}{x (1 - x)} \frac{1}{y + x
- 1} = \frac{x + y}{x y (x + y - 1)}
\end{equation}

This form could have also been derived without any triangulation.
The denominator reflects all the inequalities as it should. However,
with a random numerator, we would have non-vanishing residue at the
origin $x=y=0$, which is clearly not in the space. The numerator
kills that residue, and the resulting form has logarithmic
singularities on the boundary of our space. We could
have also arrived at this form in another way. We know the form for $x +
y < 1$. Since the form with no restriction (other than positivity)
on $x,y$ is just $1/(xy)$, we conclude that the form for $x + y > 1$
is
\begin{equation}
\frac{1}{x y } - \frac{-a}{x y (x + y - 1)} = \frac{x + y}{x y (x +
y - 1)}
\end{equation}

As a final example, let us consider $x,y,a_1,b_1,a_2,b_2 > 0$, together with the two constraints
\begin{equation}
\frac{x}{a_1} + \frac{y}{b_1} > 1, \quad \frac{x}{a_2} +
\frac{y}{b_2}
> 1
\end{equation}
We will find the form by triangulating the space in two different
ways. In the first triangulation, begin by ordering $a_1 < a_2$ without
loss of generality; the final form will be obtained by symmetrizing
$1 \leftrightarrow 2$. The shape of the allowed region $x,y$ space
depends on whether $b_1 < b_2$ or $b_1 > b_2$:
$$
\includegraphics[scale=.83]{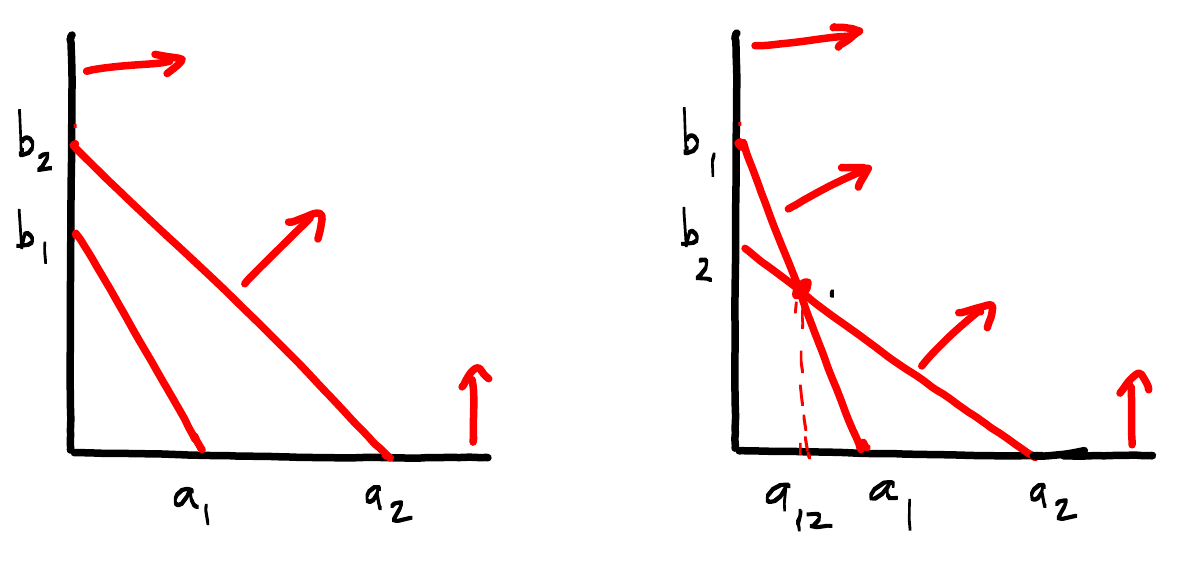}
$$
If $b_1 < b_2$, then the space is essentially the same as the
quadrilateral we just studied. The associated form obtained by
breaking it up into the two regions $x> a_2$, and $0<x<a_2$, is
given by
\begin{equation}
[a_1,a_2][b_1,b_2] \left([\underline{a_2},x] \frac{1}{y} + [x, \underline{a_2}] \frac{1}{y + \frac{b_2 x}{a_2} -b_2} \right)
\end{equation}

If $b_1 > b_2$, we have a pentagonal shape. We can break this up into three regions, where $x>a_2$, $a_2 > x > a_{12}$ and $a_{12}> x > 0$. Here
$a_{12} = \frac{a_1 a_2 (b_1 - b_2)}{a_2 b_1 - a_1 b_2}$. The associated form is
\begin{equation}
[a_1,a_2] [b_2,b_1] \left([\underline{a_2},x] \frac{1}{y} + [\underline{a_{12}},x,\underline{a_2}] \frac{1}{y+ \frac{b_2 x}{a_2} - b_2} +
[x, \underline{a_{12}}] \frac{1}{y + \frac{b_1 x}{a_1} - b_1} \right)
\end{equation}

Summing these forms and symmetrizing in $1 \leftrightarrow 2$, all
the spurious poles cancel and we find for the final form
\begin{equation}
\frac{(\frac{x}{a_1} + \frac{y}{b_1})(\frac{x}{a_2} +
\frac{y}{b_2})}{x y a_1 b_1 a_2 b_2 (\frac{x}{a_1} + \frac{y}{b_1} -
1)(\frac{x}{y_2} + \frac{y}{b_2} - 1)}
\end{equation}

Note that we could also have arrived at this result in another
simpler way, by thinking of the constraints in $(a_1,b_1)$ and
$(a_2,b_2)$ spaces separately. For fixed $x,y$, if we redefine $A_i
= x/a_i$ and $B_i = y/b_i$, we just have $A_1 + B_1 > 1, A_2 + B_2 >
1$. We then get for the form
\begin{equation}
\frac{1}{x y} \times \frac{A_1 + B_1}{A_1 B_1 (A_1 + B_1 - 1)}
\times\frac{(A_2 + B_2)}{A_2 B_2 (A_2 + B_2 - 1)}
\end{equation}
which, including the trivial Jacobian factors from the change of
variables, reduces immediately to our above result obtained using
triangulation.

\section{Two Loops}

We now move on to studying the inequalities defining the
amplituhedron for four-particle scattering, starting at  two-loops,
where we just have a single mutual positivity condition to deal
with, simply
\begin{equation}
(x_1 - x_2)(z_1 - z_2) + (y_1 - y_2)(w_1 - w_2) < 0\label{twoloop}
\end{equation}
Without loss of generality we can take $x_1 < x_2$. Then we have
\begin{equation}
z_1 - z_2 > \frac{(y_1 - y_2)(w_1 - w_2)}{x_2 - x_1}
\end{equation}
If either $y_1> y_2, w_1 > w_2$ or$y_1 < y_2, w_1 < w_2$, we have
$(y_1 - y_2)(w_1 - w_2) > 0$; the form is then
\begin{equation}
[x_1, x_2] \frac{1}{z_2} \frac{1}{z_1 - z_2 - \frac{(y_1 - y_2)(w_1
- w_2)}{x_2 - x_1}} \left([y_1,y_2][w_1,w_2] + [y_2,y_1][w_2,w_1]
\right)
\end{equation}
But if $y_1 < y_2, w_1 > w_2$ or $y_1> y_2, w_1 < w_2$, we have
\begin{equation}
z_2 - z_1 < - \frac{(y_1 - y_2)(w_1 - w_2)}{x_2 - x_1}
\end{equation}
Then the form is
\begin{align}
\frac{1}{x_1} \frac{1}{x_2 - x_1} \frac{1}{z_1} \left(\frac{1}{z_2} - \frac{1}{z_2 - z_1 + \frac{(y_1 - y_2)(w_1 - w_2)}{x_2
- x_1}}\right) \left([y_1,y_2] [w_2,w_1] + [y_2,y_1][w_1,w_2]
\right)
\end{align}
Finally, we just have to swap $1 \leftrightarrow 2$. The sum of
these terms is then
\begin{equation}
\frac{x_1 z_2 + x_2 z_1 + y_1 w_2 + y_2 w_1}{x_1 x_2 y_1 y_2 z_1 z_2
w_1 w_2 [(x_1 - x_2)(z_1 - z_2) + (y_1 - y_2)(w_1 -
w_2)]}\label{twoloopform}
\end{equation}

We can expand it as a sum of four terms by canceling terms in
numerator and denominator,
\begin{align}
\left(\frac{1}{x_2 y_1 y_2 z_1 w_1 w_2 [(x_1 - x_2)(z_1 - z_2) +
(y_1 - y_2)(w_1 - w_2)]} + 1 \leftrightarrow 2 \right) \nonumber\\ +
\left(\frac{1}{x_1 x_2 y_2 z_1 z_2 w_1 [(x_1 - x_2)(z_1 - z_2) +
(y_1- y_2)(w_1 - w_2)]}  + 1 \leftrightarrow 2
\right)\label{doublebox}
\end{align}
We can solve for all variables in terms of momentum twistors, finding
\begin{align}
\left[\frac{\la1234\ra^3}{\begin{array}{c}\la AB12\ra\la AB23\ra\la
AB34\ra\la ABCD\ra\\ \la CD34\ra\la CD14\ra\la CD12\ra\end{array}}+
\frac{\la1234\ra^3}{\begin{array}{c}\la AB23\ra\la AB34\ra\la
AB14\ra\la ABCD\ra\\ \la CD14\ra\la CD12\ra\la
CD23\ra\end{array}}\right] + {\rm symmetrization} \label{doublebox2}
\end{align}
where by symmetrization we mean adding another two terms where we swap
$(AB)\leftrightarrow(CD)$. The expression
(\ref{doublebox2}) is the integrand for two double boxes
$$
\includegraphics[scale=.98]{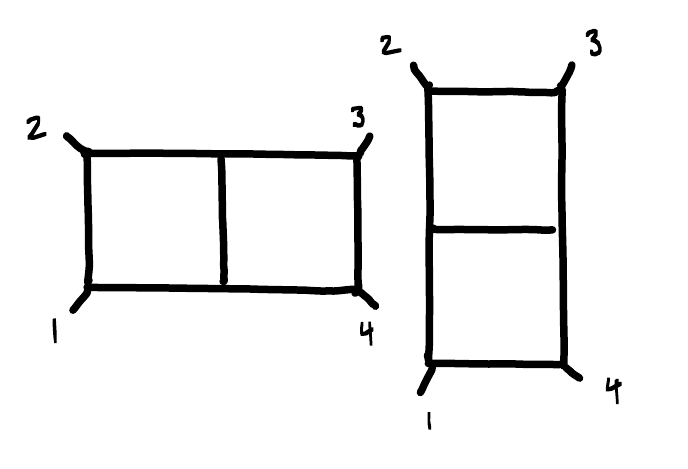}
$$
which is the standard representation of the two-loop amplitude. Note
that our approach gives the fully symmetrized (in
$(AB)\leftrightarrow(CD)$) integrand, so we get four terms
instead of two.

It is natural to ask whether some other triangulation of the space
may have directly given us this local expansion, but it is easy to
see that this is impossible: each double box individually has a cut which
is not allowed by the positivity conditions and is therefore
``outside" the amplituhedron. The cut is a simple one: suppose
we double cut one loop variable so that $D_{(1)}$ passes through the
point 1, while $D_{(2)}$ passes through the point $3$. The $D$
matrices on this cut have the form
\begin{equation}
D_{(1)} = \left(\begin{array}{cccc} 1 & 0 & 0 & 0 \\ 0 & y & 1 & z
\end{array} \right),\quad D_{(2)} = \left(\begin{array}{cccc} 1 &
x & 0 & -w \\ 0 & 0 & 1 & 0 \end{array}\right)
\end{equation}
But note that the mutual positivity condition between $D_{(1)}$ and
$D_{(3)}$ is automatically satisfied,
\begin{equation}
\langle D_{(1)} D_{(2)} \rangle = x z + y w > 0
\end{equation}
Because of this, we conclude that taking a further residue where
$(AB)_1 (AB)_2$ is cut must vanish, since there is no way to set
this to zero without further setting one of $x,z$, and $y,w$, to
zero.

This is a very simple and striking prediction of positivity, which
is true at any loop order: if we single out any two loops $(AB)_1,
(AB)_2$, and consider double cutting each so one line passes through
$1$ and the other through $3$, then the residue cutting $(AB)_1
(AB)_2$ vanishes. The vanishing of this cut is not manifest from
the local expansion. Even at two loops, each double box individually
obviously has support on this 5-cut, but the residue cancels in the
sum
$$
\includegraphics[scale=.98]{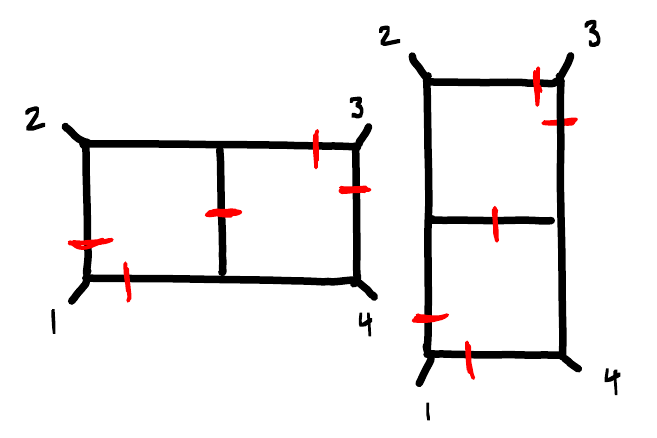}
$$
By contrast, obviously each term in our triangulation is compatible
with all positivity conditions. This mirrors familiar features of
the BCFW expansion for tree amplitudes: they correspond to
triangulations that are ``inside" the amplituhedron and manifestly
consistent with positivity properties (and therefore also the
symmetries of the theory), at the expense of manifest locality.

\section{Generalities on Cuts}

Before starting our more detailed exploration of multiloop
amplitudes, let us make some general observations about  cuts of the
integrand.

\subsection*{Reconstruction from Single Cuts}

We are familiar with reconstructing the integrand from BCFW shifts
of the external data \cite{BCFWloop}.  For instance, if we shift $Z_1 \to \hat{Z}_1 =
Z_1 + \alpha Z_4$, the integrand at $\alpha = 0$ is (the negative
of) the residues of the single cuts where $\langle (AB)_i\,\hat{1} 2
\rangle \to 0$. This is trivially reflected from positivity.  We can
divide the $w_i$ space into the pieces where $w_1$ is smallest,
$w_2$ is smallest and so on. Suppose $w_1$ is smallest; then we can
set $w_i = w_1 + \hat{w}_i$. The remaining positivity conditions are
then exactly the same as the same as the computation of the single
cut where $w_1 \to 0$. And we have to sum over the single cuts for
setting each of the $w_i \to 0$.

Obviously, we can extend this to all the variables $(x,y,z,w)$. We
can always take sum $x_{i_x}, y_{i_y}, z_{i_z},w_{i_w}$ to be
smallest, and sum over all possible $i_x,i_y,i_z,i_w$. Then, we can
compute the integrand directly by summing over all these 4-cuts.
This naturally corresponds to using a residue theorem using an
extended BCFW deformation under which $Z_1 \to \hat{Z}_1 = Z_1 + \alpha
Z_4 + \beta Z_2, \, Z_3 \to \hat{Z}_3 = Z_3 + \gamma Z_2 + \rho
Z_4$.

\subsection*{Emergent Planarity and Leading Singularities}
Let us now consider the opposite extreme, and look at the
zero-dimensional faces of the amplituhedron. Here, each $D_{(i)}$ is
taken to be one of the zero dimensional cells of $G(2,4)$, where the
columns $i,j$ can be set to the identity and the remaining entries
are zero. From the mutual positivity of equation (\ref{twoloop}), we
learn something very simple right away: the configuration will
satisfy positivity in all cases except one: we can't have the $(13)$
cells and $(24)$ cells at the same time. It is trivial to see that
this fact extends to all $n$ MHV amplitudes at all loop
orders. If all the $AB_i = (ab)_i$ are drawn as chords on a disk,
then a configuration with lines that don't cross is allowed, but
a configuration with lines that cross violates positivity and must
have vanishing residue. Examples of allowed and non-allowed
configurations are shown below:
$$
\includegraphics[scale=.63]{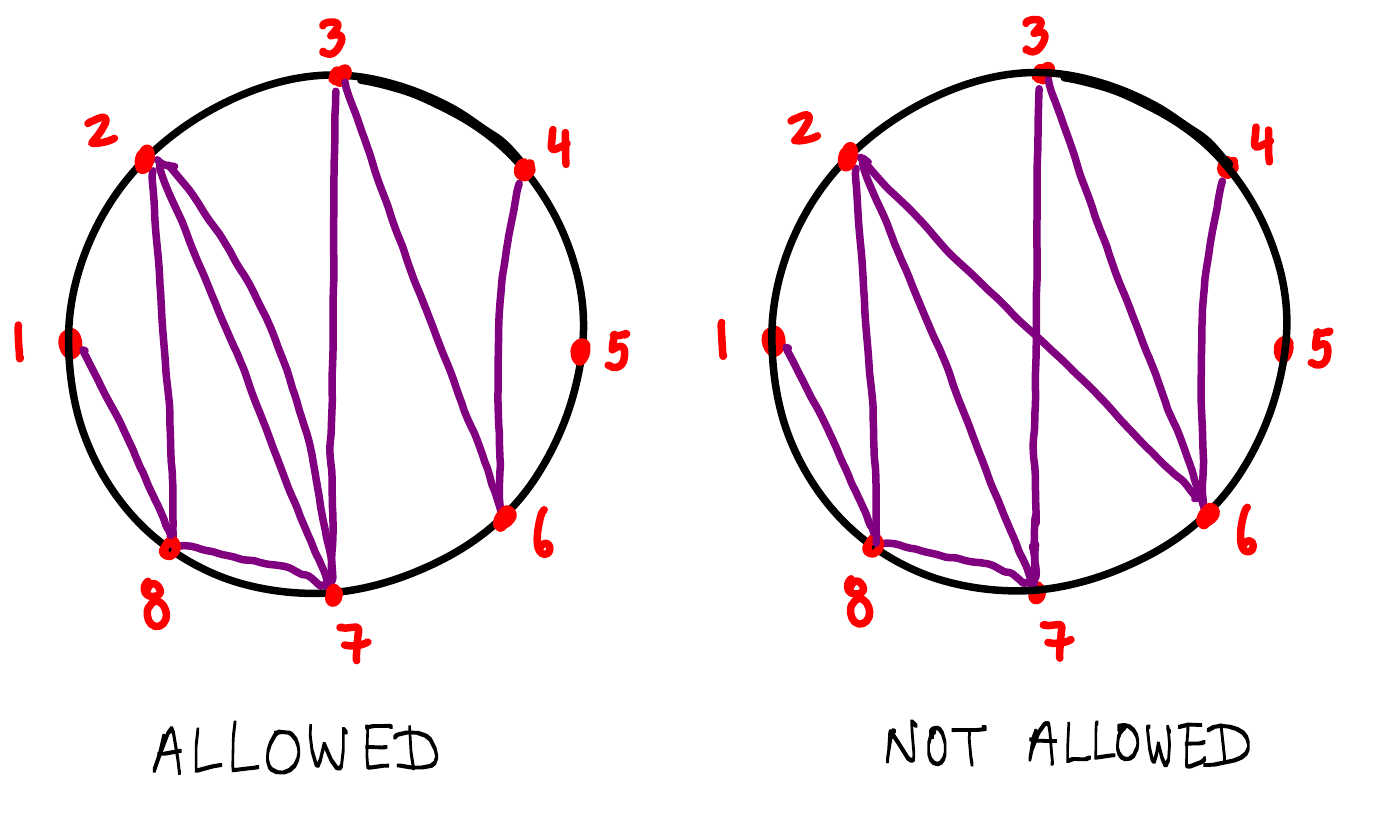}
$$

The fact that our form can ultimately be expressed as a sum over
planar local integrands is not obviously built into the
geometry, but of course does emerge from it. We see this planarity
very vividly in the above structure of leading
singularities--clearly planar diagrams can only give us leading
singularities of the allowed type, while all other objects can give
us the illegal ``crossing" configurations. Indeed there are many
meaningful local integrands, compatible with the cyclic structure on
external data, which can nonetheless not be considered as ``planar".
A simple example is the square of one-loop amplitude, whose
integrand can be written in momentum twistor space as
\begin{equation}
\frac{\la1234\ra^4}{\la AB12\ra\la AB23\ra\la AB34\ra\la AB14\ra\la
CD12\ra\la CD23\ra\la CD34\ra\la CD14\ra}
\end{equation}
This integrand has an obvious leading singularity where e.g. $AB =
13$ and $CD = 24$, which cross in index space. This is a ``not
allowed" leading singularity that is incompatible with positivity.
Thus, we see that the planar structure of the integrand is not a
trivial consequence of cyclically ordered external data, but
actually emerges from the positive geometry of the amplituhedron. Note that
``planarity"  is not an obvious invariant
property of the full integrand, but is only a natural statement
about a particular expansion of the integrand in terms of (local)
Feynman diagrams. It is thus perhaps not surprising that planarity
should be one of many derived properties of the integrand from the
amplituhedron point of view.

\section{Unitarity from Positivity}

In much of the recent work on scattering amplitudes in planar ${\cal
N}=4$ SYM, the unitarity of loop amplitudes has been directly
associated with correctly matching the single cut of the loop
integrand \cite{BCFWloop}, determined by the the forward limit \cite{Simon} of the lower-loop
amplitude. However there is an even simpler manifestation of
unitarity, familiar from the textbooks, in the double-cut (or
``unitarity cut") of the integrand, which is given by sewing
together two lower loop integrand.
$$
\includegraphics[scale=.75]{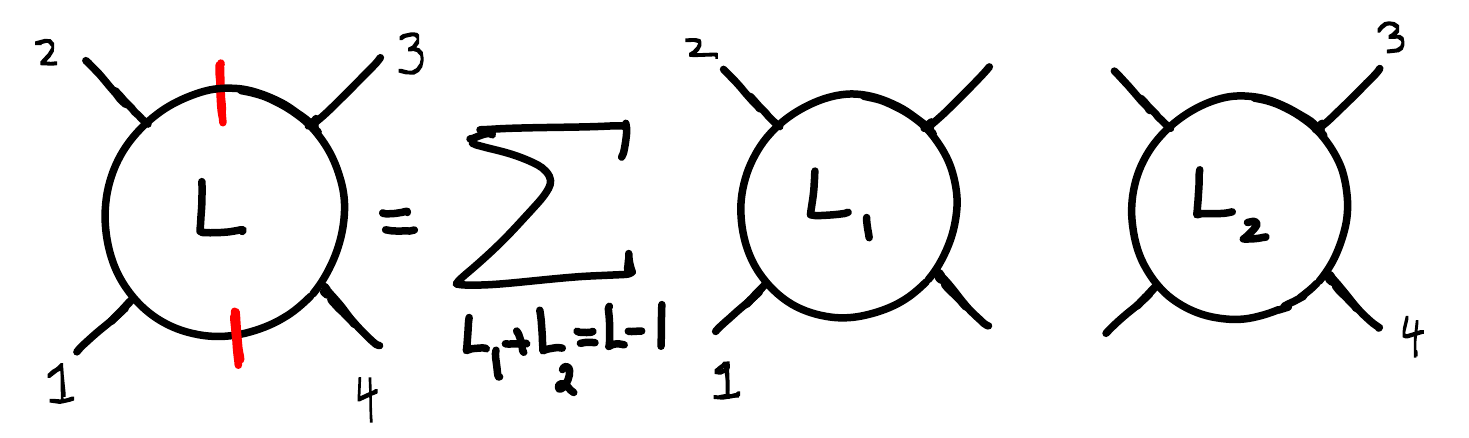}
$$
This is easy to translate to momentum-twistor language. Starting
with the $L$-loop integrand,  we take  one loop variable, $(AB)_L$,
to cut $(12)$ and $(34)$. The corresponding $D$ matrix is then of the form
\begin{equation}
D_L = \left(
        \begin{array}{cccc}
          1 & x & 0 & 0 \\
          0 & 0 & y & 1 \\
        \end{array}
      \right)
\end{equation}
If we compute the residue of the integrand on this configuration,
unitarity tells us that the result must be
\begin{equation}
\frac{dx}{x} \frac{dy}{y} \times \hspace{-0.2cm}\sum_{L_1+L_2=L-1}
M_4^{L_1}(Z_1-xZ_2,Z_2,Z_3,Z_4-yZ_3)\,\,
M_4^{L_2}\left(Z_1,Z_2-\frac{1}{x}Z_1,Z_3-\frac{1}{y}Z_4,Z_4\right)
\end{equation}
We will now see that this result follows in a simple and beautiful
way from the positive geometry of the amplituhedron.

On the unitarity cut, the positivity conditions are the usual ones
for the $(L-1)$ loop variables. For $D_L$ we just have that $x,y >
0$. The mutual positivity between $D_L$ and the remaining $D_i$ just
tells us that
\begin{equation}
(23)_i + xy(14)_i - x (13)_i - y(24)_i > 0\label{eq1}
\end{equation}
This condition also tells us that
\begin{align}
&[(13) - y(14)][(24) - x(14)] =(13)(24) - x(13)(14) - y(14)(24) + x y (14)^2 \\
&\hspace{3cm}= (12)(34) + (14)[(23) + x y (14) - x(13) - y(24)]
> (12)(34) > 0\nonumber
\end{align}
where in the second line we used $(13)(24) = (12)(34) + (23)(14)$.
Now, obviously we can divide the space of each $D_i$ into ones where
$(13)_i - (14)_i> 0$, and $(13)_i - y (14)_i< 0$, similarly $(24)_i
- x(14)_i > 0$ or $(24)_i - x (14)_i <0$. However, if the product of
these two factors is negative it is impossible to satisfy equation
(\ref{eq1}). Thus, for each $i$, we have {\it either} that
\begin{equation}
(13) - y (14) > 0 \quad {\rm and} \quad (24) - x (14) > 0
\end{equation}
or
\begin{equation}
(13) - y (14) > 0, \quad {\rm and} \quad (24) - x (14) > 0
\end{equation}

Let us say that $L_1$ of the lines $D_{a}$ satisfy the first
inequality and the remaining $L_2 = L - 1 - L_1$ lines $D_{A}$
satisfy the second inequality.  Explicitly, in the first case the
region is represented by positivity conditions
$$
(12)_a>0,(13)_a-y(14)_a>0,(14)_a>0,(23)_a>0,(24)_a-x(14)_a>0,(34)_a>0
$$
\begin{equation}
(23)_a + xy(14)_a - x (13)_a - y(24)_a> 0
\end{equation}
Let us define shifted columns
\begin{equation}
(\hat{3})_a = (3)_a- y(4)_a,\qquad (\hat{2})_a = (2)_a-x(1)_a
\end{equation}
Thus the set of positivity conditions become
\begin{equation}
(1\hat{2})_a>0,(1\hat{3})_a>0,(14)_a>0,(\hat{2}\hat{3})_a>0,(\hat{2}4)_a>0,(\hat{3}4)_a>0
\end{equation}
In the second region we have
$$
(12)_A>0,y(14)_A-(13)_A>0,(14)_A>0,(23)_A>0,x(14)_A-(24)_A>0,(34)_A>0
$$
\begin{equation}
(23)_A + xy(14)_A - x (13)_A - y(24)_A > 0
\end{equation}
Let us define shifts
\begin{equation}
(\hat{1})_A = (1)_A - \frac{1}{x}(2)_A,\qquad (\hat{4})_A =
(4)_A-\frac{1}{y}(3)_A
\end{equation}
Then the set of positivity conditions become
\begin{equation}
(\hat{1}2)_A>0,(\hat{1}3)_A>0,(\hat{1}\hat{4})_A>0,(23)_A>0,(2\hat{4})_A>0,(3\hat{4})_A>0
\end{equation}

Now, we come to the positivity conditions internal to the $D_{a}$'s, internal to the $D_{A}$'s, and also the ones between $D_{a}$ and $D_{A}$'s. Actually quite strikingly, the $D_a$'s and $D_A$'s are
automatically mutually positive! We look at
\begin{equation}
(12)_a(34)_A+(23)_a(14)_A+(34)_a(12)_A+(14)_a(23)_A-(13)_a(24)_A-(24)_a(13)_A \label{pos4}
\end{equation}
Rewriting this in terms of the natural shifted variables
\begin{equation}
(2)_a =(\hat{2})_a + x(1)_a,\quad (3)_a = (\hat{3})_a+y(4)_a \quad
(1)_A = (\hat{1})_A + \frac{1}{x}(2)_A,\quad (4)_A=
(\hat{4})_A+\frac{1}{y}(3)_A
\end{equation}
and plugging  into (\ref{pos4}) we find
\begin{align}
&(1\hat{2})_a(3\hat{4})_A+[(\hat{2}\hat{3})_a+xy(14)_a+y(\hat{2}4)_a+x(1\hat{3})_a]
\left[\frac{1}{xy}(23)_A+(\hat{1}\hat{4})_A+\frac{1}{x}(2\hat{4})_A+\frac{1}{y}(\hat{1}3)_A\right]\nonumber\\
&\hspace{-0.7cm}+(\hat{3}4)_a(\hat{1}2)_A+(14)_a(23)_A -
[(1\hat{3})_a+y(14)_a]\left[(2\hat{4})_A+\frac{1}{y}(23)_A\right]-[(\hat{2}4)_a+x(14)_a]\left[(\hat{1}3)_A+\frac{1}{x}(23)_A\right]\nonumber\\
&=(1\hat{2})_a(3\hat{4})_A+(\hat{3}4)_a(\hat{1}2)_A+\frac{1}{y}[(\hat{2}\hat{3})_a+x(1\hat{3})_a](\hat{1}3)_A
+x[(1\hat{3})_a+y(14)_a](\hat{1}\hat{4})_A\nonumber\\
&+[(\hat{2}\hat{3})_a+y(\hat{2}4)_i](14)_A+\frac{1}{x}[(\hat{2}\hat{3})_a+y(\hat{2}4)_i](2\hat{4})_A
+\frac{1}{xy}(\hat{2}\hat{3})_a(23)_A>0
\end{align}

The positivity here is quite non-trivial; the expression many terms
with plus and minus signs that cancel each other, leaving only
pluses.

The mutual positivity internally for the $D_a$'s (or the $D_A$'s) are
exactly the same for the shifted and unshifted columns, since the
$(4 \times 4)$ determinants are unchanged in shifting a column by a
multiple of another. These can be easily translated in shifts of
external twistors. Under ${\cal A}_\gamma = D_{\gamma a} \cdot Z_a$,
we have for the first shift
\begin{align}
A = D\cdot Z &= (1)Z_1+(\hat{2})Z_2+(\hat{3})Z_3
+(4)Z_4\nonumber\\&=
(1)Z_1+(2)Z_2-x(1)Z_2+(3)Z_3-y(4)Z_3+(4)Z_4\nonumber\\ &=
(1)\hat{Z_1}+(2)Z_2+(3)Z_3+(4)\hat{Z_4}
\end{align}
Thus, the form for the $L_1$ lines is
\begin{equation}
M_4^{L_1}(Z_1-xZ_2,Z_2,Z_3,Z_4-yZ_3)
\end{equation}
and analogously the form for the $L_2$ lines is
\begin{equation}
M_4^{L_2}\left(Z_1,Z_2-\frac{1}{x}Z_1,Z_3-\frac{1}{y}Z_4,Z_4\right)
\end{equation}

Thus, we conclude that the unitarity cut is
\begin{equation}
\frac{dx}{x} \frac{dy}{y} \times\hspace{-0.3cm} \sum_{L_1+L_2=L-1}
M_4^{L_1}(Z_1-xZ_2,Z_2,Z_3,Z_4-yZ_3)\,\,
M_4^{L_2}\left(Z_1,Z_2-\frac{1}{x}Z_1,Z_3-\frac{1}{y}Z_4,Z_4\right)
\end{equation}
precisely as needed to enforce unitarity.

\section{Three Loops}
Having established these general results, let us turn to the three-loop
amplitude. Recall from our general discussion  that it
suffices to look at various cuts of the amplitude, coming from
taking $x_{\sigma_1}, y_{\sigma_2}, z_{\sigma_3}, w_{\sigma_4}$ to
be smallest. For the case of three loops, at least one pair of
$\sigma_1, \sigma_2, \sigma_3, \sigma_4$ will correspond to the same
loop, thus, to compute the full three-loop integrand, it suffices to
compute the cut of the integrand where one loop is double-cut. We
have already verified that the unitarity double-cut is correctly
reproduced at any loop order. It thus suffices to compute the
remaining double cuts, which we call the ``corner cuts": where the
line passes through one of the points $Z_i$, or its parity
conjugate, where the line lies in the plane $(Z_{i-1} Z_i Z_{i+1})$.
Since these are parity conjugate, it is enough to compute one of
them, which we take to be the cut where the line corresponding to
the third loop passes through point 4. It will be convenient to use
a different gauge-fixing for the third loop
\begin{equation}
D_{(3)} = \left(\begin{array}{cccc} a & 1 & b & 0 \\ 0 & 0 & 0 & 1
\end{array} \right)
\end{equation}
If we further rescale the variables for the remaining loop variables
as $w_i \to w_i/b, y_i \to b y_i; z_i \to z_i/a, x_i \to a x_i$, the
remaining positivity conditions become
\begin{equation}
x_i + y_i > 1, \quad (x_1 - x_2)(z_1 - z_2) + (y_1 - y_2)(w_1 - w_2)
< 0
\end{equation}
We can assume that $x_1 < x_2$, so that just as for two-loops we
have then sum at the end over $1 \leftrightarrow 2$. Let us also
define
\begin{equation}
Z_+ = \frac{1}{z_2} \frac{1}{z_1 - z_2 - \frac{(y_1 - y_2)(w_1 -
w_2)}{x_2 - x_1}}, \quad Z_- = \frac{1}{z_1} \left(\frac{1}{z_2} -
\frac{1}{z_2 - z_1 - \frac{(y_2 - y_1)(w_1 - w_2)}{x_2 - x_1}}
\right)
\end{equation}
Then, by dividing the space into pieces much as we did at 2-loops,
we find that the form is
\begin{align}
&[x_1, x_2, \underline{1}] ( [\underline{1 - x_1}, y_1, y_2]
([w_1,w_2] Z_+ + [w_2,w_1]
Z_-)\\
&+([\underline{1 - x_2}, y_2, \underline{1 - x_1}, y_1] + [
\underline{1 - x_1},y_2, y_1]) ([w_2,w_1] Z_+ + [w_1,w_2]
Z_-))\nonumber\\
&+ [x_1, \underline{1}, x_2]( [\underline{1 - x_1}, y_1, y_2] ([w_1,
w_2] Z_+ + [w_2,w_1] Z_-)\nonumber\\
&+ ([y_2, \underline{1 - x_1}, y_1] + [\underline{1 - x_1}, y_2,
y_1])([w_2,w_1] Z_+ + [w_1,w_2] Z_-))\nonumber\\
&+ [\underline{1}, x_1, x_2] ([y_1, y_2]([w_1,w_2] Z_+ + [w_2,w_1]
Z_-) + [y_2,y_1]([w_2,w_1] Z_+ + [w_1,w_2] Z_-)\nonumber
\end{align}
Adding $1 \leftrightarrow 2$, all spurious poles cancel and we
obtain
\begin{align} \frac{\Bigg\{
\begin{array}{c}
w_2 x_1 x_2 y_1 + w_2 x_2 y_1^2 + w_1 x_1 x_2 y_2 - w_1 y_1 y_2 -
w_2 y_1 y_2 + w_2 x_1 y_1 y_2 + w_1 x_2 y_1 y_2 \\
+ w_2 y_1^2 y_2 +w_1 x_1 y_2^2 + w_1 y_1 y_2^2 - x_1 x_2 z_1 + x_1
x_2^2 z_1 + x_2^2 y_1 z_1 + x_1 x_2 y_2 z_1 \\
+ x_2 y_1 y_2 z_1 - x_1 x_2 z_2 + x_1^2 x_2 z_2+ x_1 x_2 y_1 z_2 +
x_1^2 y_2 z_2 + x_1 y_1 y_2 z_2
\end{array}\Bigg\}}{abx_1 x_2 y_1 y_2 z_1 z_2 w_1 w_2 (x_1 + y_1 - 1) (x_2 + y_2 - 1)
((x_2 - x_1)(z_1 - z_2) + (y_2 - y_1)(w_1 - w_2))}
\end{align}
This matches what we get from the familiar local expansion, as a sum
over ladders and ``tennis court" diagrams:
$$
\includegraphics[scale=.9]{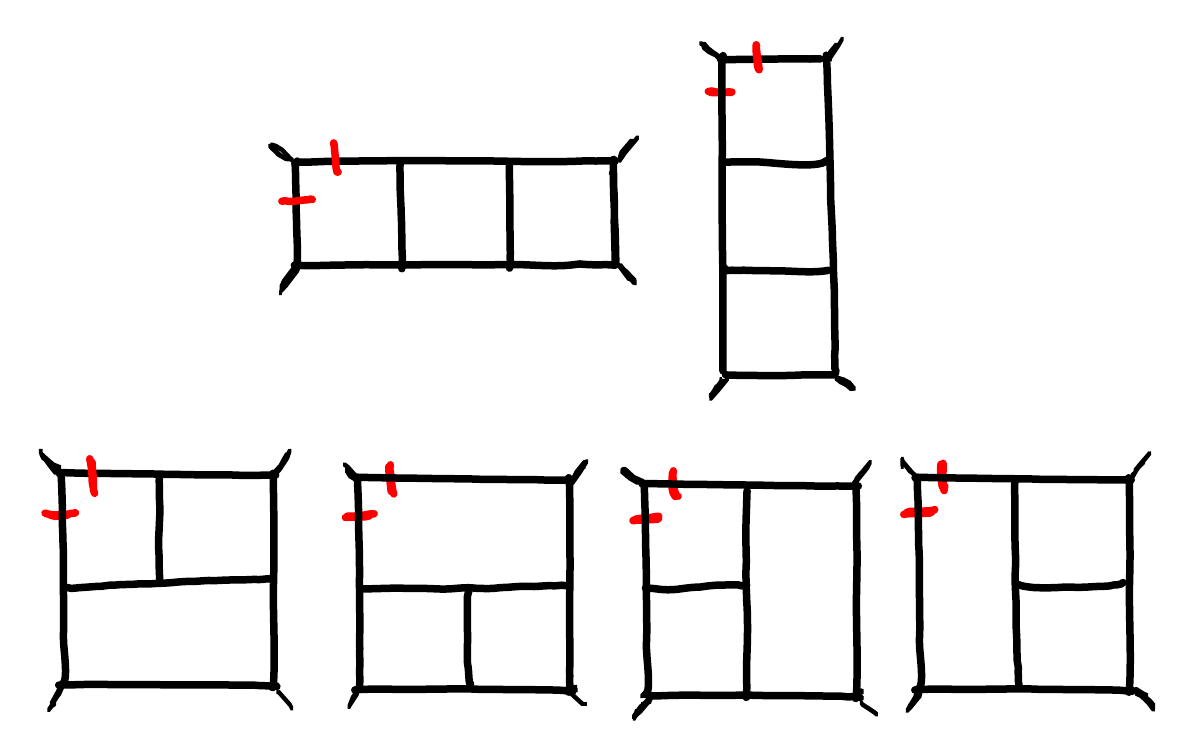}
$$

\section{Multi-Collinear Region}

We have seen that a particular double cut of a single loop--the unitarity cut-- is
simply expressed in terms of (shifted) lower-loop objects. It is thus
natural to look at the other two kinds of double cuts. Let us
consider the cut where the $L^{th}$ line passes through 2. It will
be convenient to use a different gauge-fixing for this last line
\begin{equation}
D_{(L)} = \left( \begin{array}{cccc}0 & 1 & 0 & 0 \\ -\alpha & 0 & 1
& \gamma \end{array} \right)
\end{equation}
Then the mutual positivity conditions between $D_{(L)}$ and the
other lines is simply
\begin{equation}
\alpha w_i + z_i > \gamma
\end{equation}
It is amusing that from the point of view of the
lower-loop problem, we are simply putting a simple additional
restriction on the allowed region for $w_i,z_i$.

Despite the apparent simplicity of this deformation of the $L-1$
loop problem, unlike the unitarity cut, this double cut can't be
determined in terms of shifts of lower-loop problems in a
straightforward way. However, there is a further, triple cut, which
does have a very simple interpretation. Consider the limit where
$\beta \to 0$. This is the collinear region, where the line passes
through the point 2 while lying in the plane (123) \cite{localintegrand}. Note that the
positivity condition is now automatically satisfied, and so the cut
is trivial:
\begin{equation}
A_L^{{\rm coll.}} = \frac{d \alpha}{\alpha} \times A_{L-1}
\end{equation}

In this discussion we assumed that all the lines but one are
generic. We now investigate what happens when $l$ lines are sent
into the collinear region. The most general way this can happen is
to start with $L_{12}$ lines cutting $(12)$ and $L_{23}$ lines
cutting $(23)$. Let us gauge-fix in a convenient way, and write for
the two sets of lines
\begin{equation}
D_{(i)} = \left(\begin{array}{cccc} \beta_i & 1 & 0 & 0 \\ -\alpha_i
& 0 & 1 & \gamma_i \end{array} \right) \quad D_{(I)} =
\left(\begin{array}{cccc} 0 & 1 & \rho_I & 0
\\ \alpha_I & 0 & 1 & \gamma_I \end{array} \right)
\end{equation}

In order to reach the collinear limit, we must send $\beta_i,
\gamma_i \to 0$, and $\rho_I, \delta_I \to 0$. We can send these to
zero in different ways, but let us focus on one for definiteness,
the other cases can be treated similarly. For the lines cutting
$(12)$, we first take them to pass through $2$, and then move them
into the collinear region where they lie in $(123)$; in other words,
we first send $\beta_i \to 0$, and then $\gamma_i \to 0$. Similarly
for the lines intersecting $(23)$, we first send them to pass
through $2$, then to lie in $(123)$, so that we put the $\rho_I \to
0$, then send $\gamma_I \to 0$. Now, the positivity conditions
between these lines are just
\begin{align}
(\beta_i -\beta_j)(\gamma_i - \gamma_j) > 0, \quad (\rho_I -
\rho_J)\left(\frac{\gamma_I}{\alpha_I} -
\frac{\gamma_J}{\alpha_J}\right) > 0, \quad (\beta_i - \alpha_I
\rho_I) (\gamma_i - \gamma_I) > 0
\end{align}

Collectively, these tell us something simple. Suppose we take the
lines to pass through $2$ in some particular order, say by first
taking $\beta_ 1 \to 0$, then $\beta_2 \to 0$, then $\rho_1 \to 0$,
then $\rho_2 \to 0$, then $\beta_3 \to 0$ etc. Then, the cut
vanishes unless the lines are taken into the collinear limit in
exactly the same order! In this case, the cut is just
\begin{equation}
\prod_{a=1}^{l} \frac{d \alpha_a}{\alpha_a} \times M^{L-l}
\end{equation}

\section{Log of the Amplitude}

Scattering amplitudes have well-known double-logarithmic infrared
divergences, arising precisely from loop integration in the
collinear region. At $L$ loops, we have a log$^{2L}$ divergence,
which exponentiates in a well-known way; the logarithm of the
amplitude only has a log$^2$ divergence. This is a
motivation for looking at the log of the amplitude from a
physical point of view. But as we have just seen, the loop integrand
form also has an extremely simple behavior in the multicollinear
limit. We will now see that this behavior, together with some very
simple combinatorics, already motivates looking at the logarithm of
the amplitude directly at the level of the integrand. While the
amplitude itself has a non-vanishing residue when one loop momentum
is brought into the collinear region, we will see that the log of
the amplitude vanishes in the multicollinear region, unless {\it
all} $L$ loop momenta are taken into the collinear region. The
residue depends in a non-trivial way on the specific path taken into
the collinear region. Furthermore, we will see that the log of the
amplitude naturally leads us to consider {\it all} the natural
``positive regions" we can think of related to amplituhedron
geometry.

Let us start by introducing a generating function combining together
the amplitude at all loop order, otherwise known as the amplitude
itself:
\begin{equation}
M = 1 + g M_1 + g^2 M_2 + \cdots
\end{equation}
Now, consider for any function $f$, the expansion for $f(M)$.
Suppose that
\begin{equation}
f(1 + x) = x + a_2 x^2 + a_3 x^3 + \cdots
\end{equation}
then
\begin{eqnarray}
f(M) &=& (g M_1 + g^2 M_2 + \cdots) + a_2(g^2 A_1^2 + 2 g^3 M_1 M_2
+ \cdots) + a_3 g^3 M_3 + \cdots \nonumber \\ &=& g M_1 + g^2(M_2 +
a_2 M_1^2) + g^3(M_3 + 2 a_2 M_1 M_2 + a_3M_1^3) + \cdots
\end{eqnarray}

We'd now like to extract the permutation-invariant integrand from
this expression at $L$ loops. For instance,
\begin{eqnarray}
M_3 &=& \int d^4 x_1 d^4 x_2 d^4 x_3\,\, M_3(x_1,x_2,x_3) \nonumber \\
M_1 M_2 &=& \int d^4 x_1 d^4 x_2 d^4 x_3\,\, \left[(M_1(x_1) M_2(x_2,x_3) + M_1(x_2) M_2(x_1,x_3) + M_1(x_3) M_2(x_1,x_2)  \right] \nonumber \\
M_1^3 &=& \int d^4 x_1 d^4 x_2  d^4 x_3\,\, M_1(x_1)M_1(x_2)M_1(x_3)
\end{eqnarray}
Actually, for the combinatorics, the ``$\int d^4 x$" are irrelevant.
Instead, we define a generating function
\begin{equation}
M = 1 + \sum_i x_i (i) + \sum_{i<j} x_i x_j (ij) + \sum_{i<j<k} x_i
x_j x_k (ijk) + \cdots
\end{equation}
Here $``(1)"$ stands for $M_1(x_1)$, $``(134)"$ stands for
$M_3(x_1,x_3,x_4)$ and so on. Then, the integrand for $f(M)$ at $L$
loops is just the coefficient of $(x_1 \cdots x_L)$ in the expansion
of $f(A)$, or put another way
\begin{equation}
f(M)^{L-loop} = \partial_{x_1 \cdots x_L} f(M)|_{x_1 = \cdots =
x_L = 0}
\end{equation}

Obviously, if we are only interested in $L$ loops, we can truncate
the $x_i$ to just $x_1, \cdots, x_L$ if we like. Thus, explicitly,
for $L=3$, we have
\begin{align}
M = 1 + x_1 (1) + x_2 (2) + x_3(3) + x_1 x_2 (12) + x_1 x_3 (13) +
x_2 x_3 (23) + x_1 x_2 x_3 (123)
\end{align}
and foreseeing our future interest, for $f(M) =$ log$(M)$, we have
for the 3-loop log of the amplitude
\begin{equation}
({\rm log} M)^{3-loop} = (123)+ 2 (1) (2) (3)-[(1) (23) + (2)(13) +
(3)(12)] \label{log3loop}
\end{equation}

We would now like to compute the cut of $f(M)$ in the
multi-collinear limit. Suppose $L$ lines are
sent to pass through $2$ in some order $(1,\cdots, L)$. We already
know that the cut of the amplitude in the multi-collinear limit
depends on the order in which the lines are then sent to the
collinear region--indeed for the amplitude the cut vanishes unless
the lines are sent to the collinear limit in the same order
$(1,\cdots, L)$. This will not in general be true for $f(M)$.
Suppose that a set of $l$ lines are moved into the collinear limit
in some order $\sigma = \{\sigma_1, \cdots, \sigma_l\}$. For
instance for $L=3$, we could take $\sigma = \{2\}$ or $\sigma=
\{13\}$ or $\sigma = \{231\}$. Then, it is easy to see that the
multi-collinear cut of $f(M)$ can be computed as follows.

We first
play the following game, to produce a new generating function
$M^{\sigma}$: (I) if an ordered subset of $\sigma$ occurs out of
order in the brackets of $M$, we drop that term. (II) We then delete
all the labels in $\sigma$. Let's illustrate this for $L=3$ with
$\sigma = \{31\}$. The terms $x_1 x_3 (13)$ and $x_1 x_2 x_3 (123)$
in $M$ are dropped, and the rest are kept. Then, we drop the
indices $3,1$, and are left with
\begin{equation}
M^{\{31\}} = 1 + x_1 + x_2 (2) + x_3 + x_1 x_2 (2) + x_2 x_3 (2)
\end{equation}

The multi-collinear cut of $f(M)$ is easily seen to be just
\begin{equation}
\prod \frac{d \alpha_{\sigma_i}}{\alpha_{\sigma_i}} \times
\partial_{x_1, \cdots x_L} f(M^\sigma)|_{x_1 = \cdots = x_L = 0}
\end{equation}

We are now ready to see why the logarithm of the amplitude is so
natural from a purely combinatorial point of view. Let us return to
looking at $M^{\{ 31 \}}$. Observe that while $M$ is itself and
irreducible polynomial, $M^{\{31\}}$ factorizes as
\begin{equation}
M^{\{31\}} = [1+ x_1 + x_3][1 + x_2 (2)]
\end{equation}
Note that the second factor is just the $M$ polynomial made of the
undeleted variables, while the first factor is the generating
function made of the variables with only terms in correct order
kept.

This is a general statement. For any $\sigma$, let
$\bar{\sigma}$ be the complementary set. Then
\begin{equation}
M^{\sigma} = \left(1 + \sum_i x_{\sigma_i} + \sum_{i<j; \sigma_i <
\sigma_j} x_{\sigma_i} x_{\sigma_j} + \cdots\right) \times
M^{\bar{\sigma}}
\end{equation}
To illustrate with a more non-trivial example, say for $L=5$ and
$\sigma = \{415\}$, we have
\begin{align}
M^{\{415\}} = (1 + x_1 + x_4 + x_5 + x_1 x_5 + x_4 x_5) \times (1 +
x_2(2) + x_3(3) + x_2 x_3 (23))
\end{align}

Note that if $\sigma$ is anything other than the empty set, the
factorization in non-trivial. Because of this factorization, it is
natural to consider the log of the object. Then, we see that for any
string $\sigma$ of length $0 < l < L$,
\begin{equation}
\partial_{x_1\cdots x_L} {\rm log} M^{\sigma} = 0
\end{equation}

We thus learned something remarkable: if we take the log of the
amplitude, then the cut taking any number $l < L$ of the loop
variables into the collinear region vanishes! Only if all $L$ lines
are taken into the collinear region together, can we get something
non-zero. This explains why the log of the amplitude only has a
log$^2$ divergence. We get a log$^2$ divergence from each loop
momentum brought into this collinear region. For the amplitude
itself, we can bring all $L$ lines into the collinear region one at
a time, and thus we get the log$^{2L}$ IR divergence. However for
the log of the amplitude, since all $L$ lines must be brought in the
collinear region together we just get a single overall log$^2$
divergence. (Note that had even this limit given us no residue, the
log would have have been completely IR divergence free!).

Very interestingly, the logarithm of the amplitude doesn't have the
property, familiar for the amplitude itself, of having ``unit
leading singularities". If all $L$ lines are taken into the
collinear region in an order $\sigma = (\sigma_1, \cdots,
\sigma_L)$, then the residue is
\begin{equation}
\partial_{x_1 \cdots x_L} {\rm log}\left(1 + \sum_i x_i + \hspace{-0.2cm}\sum_{i<j; \sigma_i < \sigma_j} x_i x_j + \cdots\right)|_{x_i=0}
\end{equation}

Note that unlike the amplitude itself, which is only non-vanishing
for $\sigma = (1,2,\cdots, n)$, the log of the amplitude vanishes in
this case, since
\begin{equation}
\left(1 + \sum_i x_i + \hspace{-0.2cm}\sum_{i<j; \sigma_i <
\sigma_j} x_i x_j + \cdots\right) = (1 +x_1) \cdots (1+x_n)
\end{equation}
maximally factorizes! In the other extreme, if $\sigma = (n,
\cdots, 1)$ is oppositely ordered to $(1, \cdots, n)$, then we have
\begin{equation}
\partial_{x_1 \cdots x_L} (1 + x_1 + \cdots x_L)|_{x_i = 0}  = (L-1)!
\end{equation}
In general, we can find residues ranging from $1$ to $(L-1)!$. For instance, at 4 loops,
we have the non-vanishing residues 1,2,3, 4 and 6, coming from the following paths:
\begin{align}
&1: (2,3,4,1)(2,4,1,3)(3,1,4,2)(4,1,2,3)\quad
2:(2,4,3,1)(3,2,4,1)(4,1,3,2)(4,2,1,3) \nonumber\\
&3:(3,4,1,2) \quad4: (3,4,2,1)(4,2,3,1)(4,3,1,2) \quad 6:(4,3,2,1)
\quad 0:\mbox{other}
\end{align}

The log of the amplitude has another fascinating feature, which we
can see already starting at 2-loops, where
\begin{equation}
{\rm log} M_2 = (12) - (1)(2)
\end{equation}

Note that the 2-loop amplitude puts the positivity restriction
$\langle D_{(1)} D_{(2)} \rangle > 0$ on the lines, but
``$1-$loop$^2$" part does not put any positivity restrictions on
them. Indeed, we can think of this as the sum over two regions, with
$\langle D_{(1)} D_{(2)} \rangle > 0$ and $\langle D_{(1)} D_{(2)}
\rangle < 0$. Thus, the sum that gives the log is the form
associated with the region where $\langle D_{(1)} D_{(2)} \rangle <
0$ ! The pattern continues at all higher loops. At 3-loops we have
three positivity conditions involving
\begin{equation}
\{\la D_{(1)} D_{(2)}\ra,\la D_{(1)} D_{(3)}\ra,\la
D_{(2)}D_{(3)}\ra\}
\end{equation}
For the amplitude they are all positive, $M_3 =  \{+++\}$ while for
the log of the amplitude (\ref{log3loop}) we get a sum of terms
\begin{equation}
({\rm log}\,M)^{3-loop} = \{+--\} \oplus \{-+-\} \oplus \{--+\}
\oplus 2\{---\}
\end{equation}

At 4 loops we have 6 positivity conditions,
\begin{equation}
\{\la D_{(1)}D_{(2)}\ra,\la D_{(1)} D_{(3)}\ra,\la D_{(1)}D_{(4)}
\ra, \la D_{(2)}D_{(3)}\ra, \la D_{(2)}D_{(4)}\ra, \la
D_{(3)}D_{(4)}\ra\}
\end{equation}
For the amplitude we have $M_4=\{++++++\}$. The log is
\begin{align}
&({\rm log}\,M)^{4-loop}=
(1234) -[(12)(34) + (13)(24) + (14)(23)] \nonumber\\
&\hspace{1cm}+[(12)(3)(4) + (13)(2)(4) + (14)(2)(3) + (23)(1)(4) +
(24)(1)(3)+(34)(1)(2)] \nonumber\\
&\hspace{1cm}+2\,[(123)(4) + (124)(3) + (134)(2) + (234)(1)] -
6\,(1)(2)(3)(4) \label{log4}
\end{align}
and can be decomposed into a sum of regions as
\begin{equation}
({\rm log}\,M)^{4-loop} = R_1 \oplus 2R_2 \oplus 3R_3 \oplus
4R_4\oplus 6R_6
\end{equation}
where
\begin{align*}
R_1 =&
\{---+++\}\oplus\{--++-+\}\oplus\{--+++-\}\oplus\{-+--++\}\\&\oplus\{-+-++-\}\oplus\{-++--+\}\oplus\{-++-+-\}\oplus\{-+++--\}\\
&\oplus\{+---++\}\oplus\{+--+-+\}\oplus\{+-+--+\}\oplus\{+-+-+-\}\\&\oplus\{+-++--\}\oplus\{++---+\}\oplus\{++--+-\}\oplus\{++-+--\}\\
R_2=&\{----++\}\oplus\{---+-+\}\oplus\{---++-\}\oplus\{--+--+\}\\&\oplus\{--+-+-\}\oplus\{-+---+\}\oplus\{-+-+--\}\oplus\{-++---\}\\
&\oplus\{+---+-\}\oplus\{+--+--\}\oplus\{+-+---\}\oplus\{++----\}\\
R_3=&\{--++--\}\oplus\{-+--+-\}\oplus\{+----+\}\\
R_4=&\{-----+\}\oplus\{----+-\}\oplus\{---+--\}\oplus\{--+---\}\\
&\oplus\{-+----\}\oplus\{+-----\}\\
R_6= &\{------\}
\end{align*}
While the expansion of the logarithm itself includes terms with both plus and minus signs,
remarkably, in all cases we get a sum over regions, with all
positive integer coefficients, reflecting the allowed leading
singularities for different orderings of approaching the collinear
region.

\newpage

\section{Some Faces of the Amplituhedron}

In this section, we study a few classes of lower-dimensional faces
of the amplituhedron, that are particularly easy to triangulate. The
canonical form associated with these faces computes corresponding
cuts of the full integrand.

\subsection*{Ladders and Next-to-Ladders}
Already in \cite{P1}, we discussed a set of faces that are
extremely easy to understand. Let's take all $L$ loops to cut
the line $(12)$, by sending all the $w_i \to 0$. The positivity
conditions just become $(x_i - x_j)(z_i - z_j) < 0$. In whatever
configuration of $x$'s we have, they are ordered in some way, say
$x_1 < \dots < x_L$, and this condition tells us that the $z$'s are
oppositely ordered $z_1 > \dots
> z_L$. The $y_i$ just have to be positive. The associated form is
then trivially
\begin{equation}
\frac{1}{y_1} \dots \frac{1}{y_L} \frac{1}{x_1} \frac{1}{x_2 - x_1}
\dots \frac{1}{x_L - x_{L-1}} \frac{1}{z_L} \frac{1}{z_{L-1} - z_L}
\dots \frac{1}{z_1 - z_2}
\end{equation}
which corresponds to the unique ``ladder" local diagrams that can
contribute to this cut; to see this propagator structure explicitly,
we simply regroup the terms in the product as $1/(y_1 \cdots y_L)$
multiplying
\begin{equation}
\frac{1}{x_1} \times \frac{1}{(x_2 - x_1)(z_1 - z_2)} \times
\cdots \times \frac{1}{(x_L - x_{L-1})(z_{L-1} -z_L)} \times
\frac{1}{z_L}
\end{equation}

We can move on to consider ``next-to-ladder" cuts. Suppose for
instance that $(L-1)$ of the loop variables cutting $(12)$, while
the $L$'th loop cuts $(34)$ so that $y_L \to 0$. The positivity for
the $(L-1)$ lines is simply $x_1<x_2< \cdots < x_{L-1}$ and $z_1 >
z_2 > \cdots> z_{L-1}$ as above. The mutual positivity conditions
are just that
\begin{equation}
w_L y_i > (x_i - x_L)(z_i - z_L)
\end{equation}

The canonical form is very easy to determine. We simply consider all
the for $L$ orderings of the $x$'s for which $x_1 < \cdots, <
x_{L-1}$, i.e. the orderings $[x_1, \cdots, x_{L-1}, x_L]$, $[x_1,
\cdots, x_L, x_{L-1}]$, $\cdots$, $[x_L, x_1, \cdots, x_{L-1}]$;
similarly, we consider all the analogous orderings of the $z$'s:
$[z_L, z_{L-1},\cdots, z_1]$, $[z_{L-1},z_L,\cdots,z_1]$, $\cdots$,
$[z_{L-1}, \cdots, z_1, z_L]$, Then if in the ordering, either both
$x_k > x_L$ , $z_k > z_L$ or $x_k < x_L, z_k < z_L$, we have $y_k >
(x_k - x_L)(z_k - z_L)/w_L$, otherwise we just have $y_k > 0$. The
corresponding form is

\begin{align}
&\sum_{\sigma_1 < \cdots < \sigma_{L-1}, \rho_1
> \cdots > \rho_{L-1}} [x_{\sigma^{-1}_1}, \cdots,
x_{\sigma^{-1}_L}] [z_{\rho^{-1}_1}, \cdots, z_{\rho^{-1}_L}]\\
&\hspace{0.5cm}\times \prod_{k = 1}^{L-1} \left\{
\begin{array}{c} [y_k - \frac{1}{w_L}(x_k - x_L)(z_k - z_L)]^{-1} \,
\sigma_k > \sigma_L,
\rho_k > \rho_L \quad {\rm or}\quad \sigma_k < \sigma_L, \rho_k < \rho_L\\
y_k^{-1} \,\,\, {\rm otherwise} \end{array} \right\}\nonumber
\end{align}

This expression sums the cuts for local diagrams of the form
$$
\includegraphics[scale=.85]{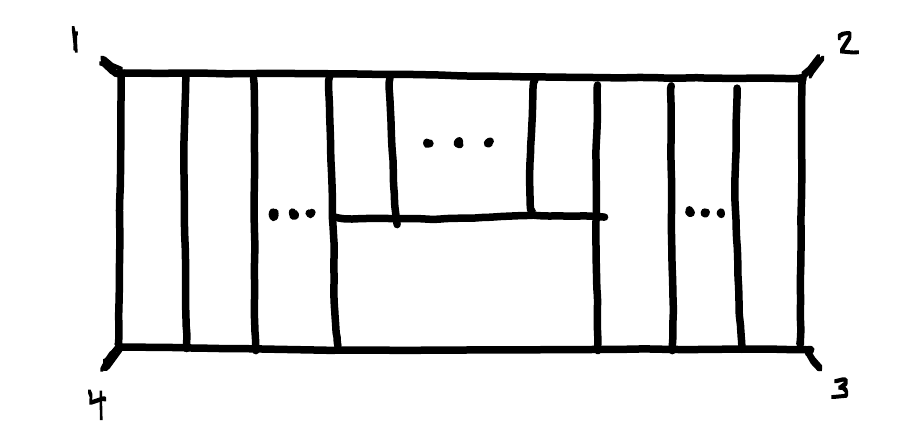}
$$

\subsection*{Corner Cuts}
We can systematically approach the faces of the amplituhedron where
every line is one of the double-cut configurations. We already know
what happens with the unitarity double-cut on general grounds. So we
are left with the ``corner cuts", where any line either passes
through $Z_i$, or lies in the plane $(Z_{i-1} Z_i Z_{i+1})$. We use
different convenient gauge fixings: for the case of lines passing
through $1$, and lines in the plane $(412)$, we use
\begin{equation}
D_{{\rm through \, 1}} = \left(\begin{array}{cccc} 1 & 0 & 0 & 0 \\
0 & y & 1 & z
\end{array} \right), \quad D_{{\rm in \, (412)}} =
\left(\begin{array}{cccc} u & 1 & 0 & 0 \\ - v & 0 & 0 & 1
\end{array} \right)
\end{equation}

Note that
\begin{equation}
\langle D_{{\rm through\, 1}} D_{{\rm in \,(412)}} \rangle = - 1
\end{equation}
is negative, and so we immediately learn that it is impossible to
have lines of both types in one corner! We can either have a
collection of lines passing through 1, {\it or} a collection of
lines lying in the plane $(412)$. Suppose we approach the
configuration where all the lines path through $1$, by starting with
all the lines intersecting $(41)$, and sending the lines into the
corner in some order, first $w_1 \to 0, \cdots,$ then $w_L > 0$.
This orders $w_1 < \cdots < w_L$ and so $y_1 > \cdots > y_L$, thus
the form on this final corner cut is just $[y_L, \cdots, y_1]$. Note
that we see again something we have observed already a number of
times: the form on the cut depends not just on the geometry of the
ultimate configuration of lines, but also on the path taken to that
configuration.

We can easily determine completely general
corner cuts where all the lines are of one type or the other. For
instance, suppose we start with $L_1$ lines cutting $(14)$, and
$L_2$ lines cutting $(12)$, and that we send these lines to pass
through the corners $1$ and $2$ in some order. If we parametrize the
matrices as
\begin{equation}
\left(\begin{array}{cccc} 1 & 0 & 0 & 0 \\ 0 & y_i & 1 & z_i
\end{array} \right) \, \left(\begin{array}{cccc} 0 & 1 & 0 & 0 \\ -
\alpha_I & 0 & \beta_I & 1 \end{array} \right)
\end{equation}
then the positivity conditions are just $y_{L_1} > \cdots > y_1,
\beta_{L_2}> \cdots > \beta_1$, with the mutual positivity condition
$z_i \beta_I > 1$, which just means $z_i > 1/\beta_1$ for all $i$.
Then the form is trivially
\begin{equation}
\prod_I \frac{1}{\alpha_I} \times [y_{L_1}, \cdots, y_1] [\beta_{L_2}, \cdots, \beta_1] \prod_i
\frac{1}{z_i - \beta_1^{-1}}
\end{equation}

This result generalizes trivially to the case with $L_1$ lines
cutting $(41)$, $L_2$ lines cutting $(12)$, $L_3$ lines cutting
$(23)$ and $L_4$ lines cutting $(34)$, then taken to pass through
$1,2,3,4$.
$$
\includegraphics[scale=.75]{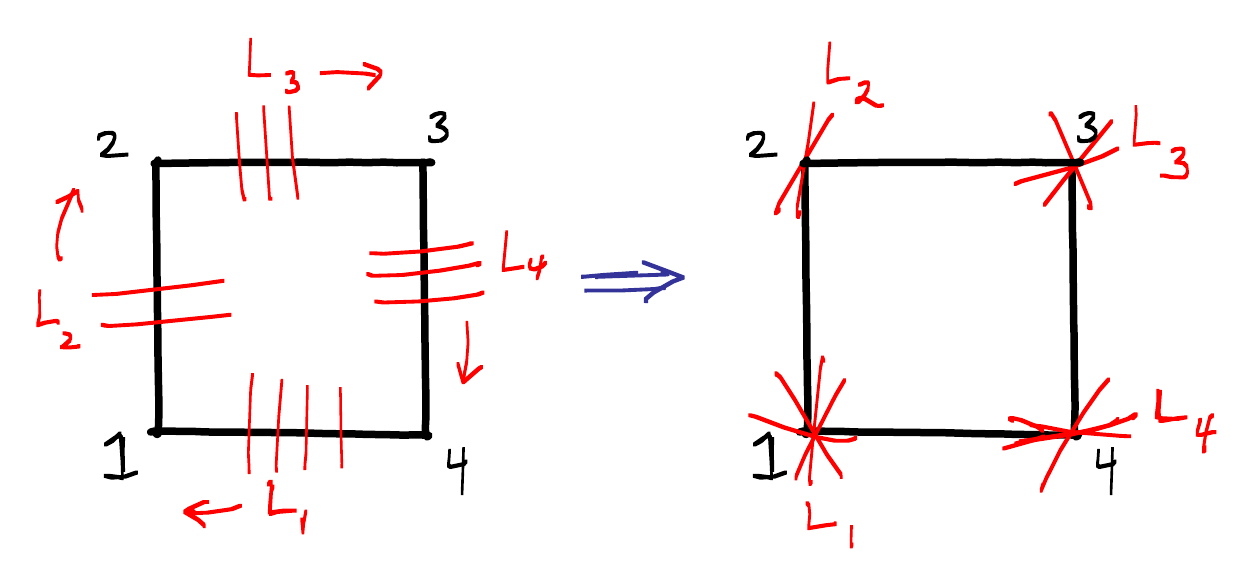}
$$

These results are very simple and arise from a single local term.
Much more interesting are the mixed corner cuts, where we have the
two different types of lines passing through different corners. One
case is still extremely simple, where the two different lines pass
through consecutive corners. Suppose we have $L_1$ lines passing
through $1$, and $L_2$ lines lying in $(123)$. It is trivial to see
that
\begin{equation}
\langle D_{{\rm through \, 1}} D_{{\rm lying \, in \, (123)}}
\rangle = (14)_{{\rm through \, 1}} (23)_{{\rm lying \, in (123)}} >
0
\end{equation}
and so the mutual positivity between these two sets is automatically
satisfied. The form is then just the product of the form for the
$L_1$ lines and the $L_2$ lines separately.

The non-trivial case is
when the corner cuts are different lines in opposite corners.
Suppose we have $L_1$ lines cutting $(41)$ that were then sent to
pass through $1$ in order $(L_1, \cdots, 1)$, and $L_3$ lines
cutting $(23)$ that were made to pass through $(234)$ in order
$(1,\cdots,L_3)$.
$$
\includegraphics[scale=.75]{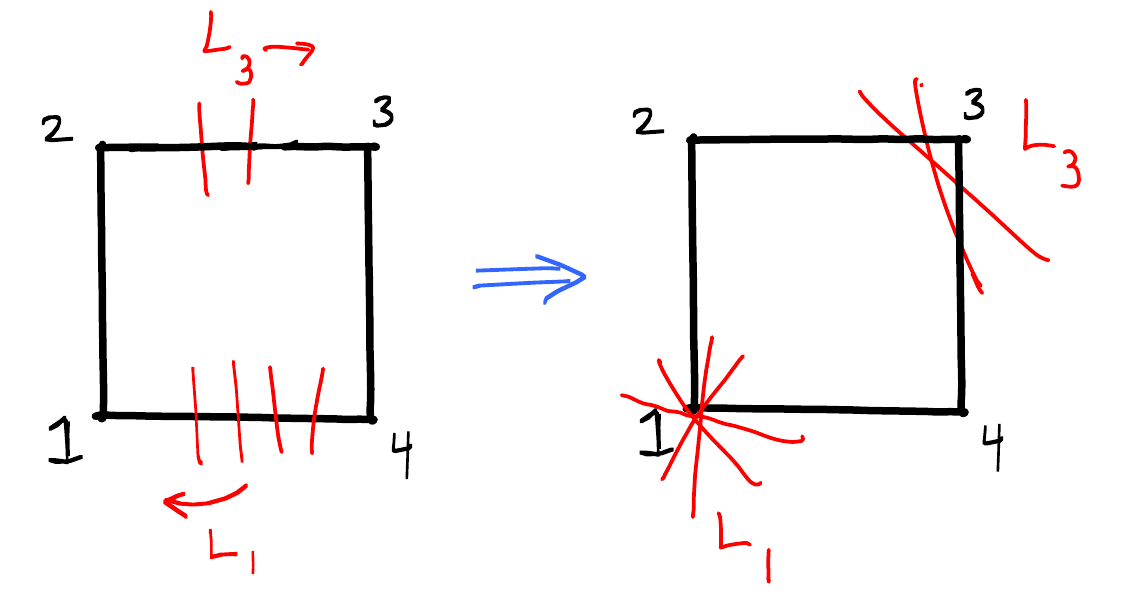}
$$
For notational convenience we'll parametrize the $D$ matrices using different
variable names in this case:
\begin{equation}
D_{{\rm through \, 1}} = \left(\begin{array}{cccc} 1 & 0 & 0 & 0
\\ 0 & x_i & 1 & y_i
\end{array} \right),\quad D_{{\rm lying \, in (234)}} =
\left(\begin{array}{cccc} 0 & 1 & a_I^{-1} & 0 \\ 0 & 0 & b_I^{-1} &
1 \end{array} \right)
\end{equation}
Then the positivity conditions are
\begin{equation}
x_1 < \cdots < x_{L_1}, \, a_1 <
\cdots < a_{L_3}, \, \, {\rm and} \, \, \frac{x_i}{a_I} +
\frac{y_i}{b_I} > 1
\end{equation}

It is quite straightforward to triangulate this space; let us work
out the case $L_3 = 2$ explicitly. Here the geometry is very similar
to the last of our warmup exercises. Suppose first that $b_1 < b_2$.
Then the inequalities are just $x_i/a_2 + y_i/b_2 > 1$ together with
the restriction $x_1 < \cdots < x_{L_1}$. We simply order the $x_i$
relative to $a_2$. If $x_i > a_2$, then we just have $y_i > 0$ and
the form is $1/y_i$, while if $x_i < a_2$, we have $Y_{i,2} > 0$ and
the form is $1/Y_{i,2}$. Here we have defined
\begin{equation}
Y_{i,1} = y_i + \frac{b_1 x_i}{a_1} - b_1, Y_{i,2} = y_i + \frac{b_2
x_i}{a_2} - b_2.
\end{equation}
Thus, for $b_1 < b_2$, the form is just
\begin{equation}
\frac{1}{a_1 (a_2 - a_1)} \frac{1}{b_1(b_2 - b_1)} \sum_m [\cdots,
x_m, \underline{a_2},x_{m+1} \cdots ] \prod_k
\left\{\begin{array}{c} Y_{k,2}^{-1} \, \, k \leq m \\ y_k^{-1} \,
\, k > m \end{array} \right\}
\end{equation}

If instead $b_2 < b_1$, then we have to break $x$ space up into the
three regions between $0,a_{12},a_2$ where $a_{12} = \frac{a_1 a_2
(b_1 - b_2)}{a_2 b_1 - a_1 b_2}$. We have to sum over all the
orderings of the $x$'s relative to $a_{12},a_2$'; for all the $x_i >
a_2$, the form in $y$ space is just $1/y_i$, for the $x_i$ in the
range $a_2 > x_i > a_{12}$ the $y$ form is just $1/Y_{i,2}$, while
for $a_{12} > x_i > 0$ the $y$ form is $1/Y_{i,1}$. Thus in this
case the form is
\begin{equation}
\frac{1}{a_1 (a_2 - a_1)} \frac{1}{b_1(b_2 - b_1)} \sum_{m \leq l}
[\cdots, x_m,\underline{a_{12}},x_{m+1}, \cdots, x_l, \underline{a_2},x_{l+1}, \cdots]
 \prod_k \left\{\begin{array}{c} Y_{k,1}^{-1}, \, \, k \leq m \\
Y_{k,2}^{-1}, \, \, m < k \leq l \\ y_k^{-1} \, \, k > l \end{array}
\right\}
\end{equation}

The full form is just the sum of these two pieces. While this result
is completely straightforward from triangulation, it gives rise to
highly non-trival local expressions even at comparatively low loop
order. In the first really interesting case at 5 loops, with $L_1 =
3$ and $L_3 = 2$, 19 local terms contribute to this cut, and when they are all combined under a common
denominator, the numerator has 325 terms.

There is another interesting feature of these cuts, that is not
evident from any traditional point of view but is obvious from the
positive geometry. We have seen that fixing the order in which the
lines are brought to pass through $1$, imposes the constraint $x_1 <
\cdots < x_{L_1}$. But, if we sum over all the different orderings,
we simply remove these ordering constraints! We then expect that the
form simplifies greatly. Indeed, if we stick with the case $L_3 =
2$, then we just get several copies of the problem $x/a_1 + y/b_1 >
1, x/a_2 + y/b_2 > 1$, which we analyzed in our warmup section.
Thus, the sum over all the ways to start with $L_1$ lines on $(41)$
which are then sent through $1$ (while sending $L_3 = 2$ lines to
lie in (234) in the usual fixed order), is
\begin{align}
\frac{1}{a_1 (a_2 - a_1)}&\left[ \frac{1}{b_1 (b_2 - b_1)} \prod_i
\left([x_i, \underline{a_2}] \frac{1}{Y_{i,2}} +
[\underline{a_2},x_i] \frac{1}{y_i}\right)\right.\\
&\hspace{0.3cm}\left.+ \frac{1}{b_2 (b_1 - b_2)} \prod_i \left([x_i,
\underline{a_{12}}] \frac{1}{Y_{i,1}} + [\underline{a_{12}}, x_i,
\underline{a_2}] \frac{1}{Y_{i,2}} + [\underline{a_2},x_i]
\frac{1}{y_i}\right) \right]\nonumber
\end{align}

\subsection*{Internal Cuts}

It is interesting that up to 4 loop order, every loop in the local
expansion of the amplitude touches the external lines, but this
behavior is obviously not generic. Starting at 5 loops, we have
diagrams with purely internal loops, such as
$$
\includegraphics[scale=.8]{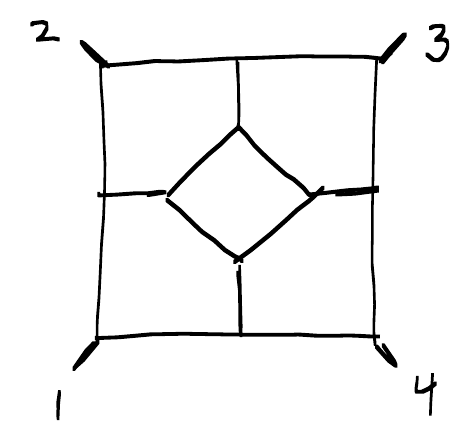}
$$
and it is interesting to probe these from positivity. Let us look at
a particularly simple set of cuts that exposes the structure in a
nice way. Suppose we take 4 lines $(AB)_1, \cdots, (AB)_4$, and take
them to pass through $1,2,3,4$ respectively. But additionally, we
take the cut where $\langle AB_1 AB_2 \rangle \to 0$, $\langle AB_2
AB_3 \rangle \to 0$, $\langle AB_3 AB_4 \rangle \to 0$, $\langle
AB_4 AB_1\rangle \to 0$, i.e. the lines are taken to one intersect
the next. The $D$ matrices
are simply
\begin{align}
D_{(1)} = \left(\begin{array}{cccc} 1 & 0 & 0 & 0 \\
0 & \alpha^{-1} & 1 & \beta \end{array} \right), \quad D_{(2)} =
\left(\begin{array}{cccc} 0 & 1 & 0 & 0 \\ \gamma & 0 & \beta^{-1} &
1 \end{array} \right),\nonumber\\ D_{(3)} =
\left(\begin{array}{cccc} 0 & 0 & 1 & 0 \\ -1 & -\sigma & 0 &
\gamma^{-1} \end{array} \right),\quad D_{(4)} =
\left(\begin{array}{cccc} 0 & 0 & 0 & 1 \\ - \sigma^{-1} & -1 & -
\alpha & 0 \end{array} \right)
\end{align}

Note that the mutual positivity between $D_{(1)} D_{(3)}$ and
$D_{(2)} D_{(4)}$ is automatic. The geometry of the lines is
$$
\includegraphics[scale=.7]{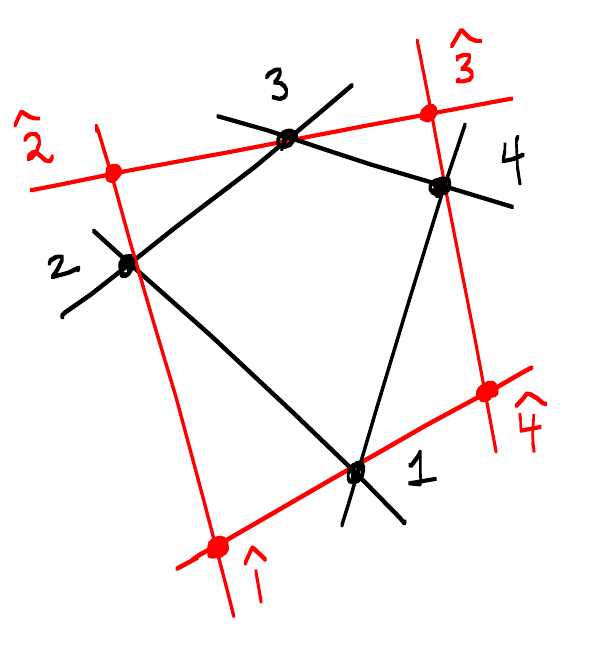}
$$
and so we can think of the lines as $(AB)_1 = \hat{1} \hat{2},
(AB)_2 = \hat{2} \hat{3}, (AB)_3 = \hat{3} \hat{4}, (AB)_4 = \hat{1}
\hat{4}$, where
\begin{align}
\hat{1} = 1 + \sigma(2 + \alpha(3 + \beta 4)), \quad \hat{2} = 2 +
\alpha(3 + \beta(4 - \gamma 1)) \nonumber\\ \hat{3} = 3 + \beta(4 -
\gamma(1 + \sigma 2)),\quad \hat{4} = 4 - \gamma(1 + \sigma(2 +
\alpha 3))
\end{align}

Now, it is easy to see that the remaining mutual positivity
conditions between $D_{(1)}, \cdots, D_{(4)}$ and the other
$D_{(i)}$ are just satisfied by the lower-loop shifted amplitude;
thus we conclude that on this cut the form is
\begin{equation}
\frac{d \alpha}{\alpha} \frac{d \beta}{\beta} \frac{d
\gamma}{\gamma} \frac{d \sigma}{\sigma} \times M^{L-4}(\hat{1},
\hat{2}, \hat{3}, \hat{4})
\end{equation}

\section{Four Particle Outlook}

We have only scratched the surface of the rich amplituhedron
geometry controlling four-particle scattering in planar ${\cal N} =
4$ SYM at all loop order. There is obviously much more to be done
just along the elementary lines of this note,  minimally in further continuing a
systematic exploration of other facets of the geometry,
corresponding to different classes of cuts of physical interest. But
we close with a few comments about some different avenues of
exploration.

In this note we have approached the determination of the integrand
for four-particle scattering  by directly ``triangulating" the
amplituhedron geometry. The $L-$ loop geometry is defined in a
self-contained way, as a subspace living inside $L$ copies of
space-time realized as $G(2,4)$. In particular, no-where do we need
to refer to lower-loop, higher-$k$ amplitudes, as in necessary in
the BCFW recursion approach \cite{BCFW2} to loop integrands
\cite{BCFWloop}. Nonetheless, it is likely that some natural
connection exists with the full problem, and perhaps a broader view
of the bigger amplituhedron geometry in which the four-particle
problem sits will be important for systematically determining the
all-loop integrand. Certainly, experience with the positive Grassmannian
\cite{alex, positive} strongly suggests that different faces  can't be properly  understood in
isolation.

As we have seen, the approach to computing the integrand by
triangulating the amplituhedron does not give us the familiar
expansions that are manifestly local. This is of course not
surprising; however, what {\it is} surprising is that some special
local expansions expose yet {\it another} aspect of positivity, that
we are not making apparent in the triangulation approach. As also
mentioned in \cite{P1}, we are still clearly missing a picture of
the form $\Omega$ which is analogous to one available for convex
polygons, determined by a literal volume of the {\it dual} polygon.
We don't yet have a notion of a ``dual amplituhedron", but there is
a powerful indication that such a formulation must exist: the form
$\Omega$ is itself positive, inside the amplituhedron! More
specifically, we can write the $L-$loop integrand as
\begin{equation}
\Omega_L(AB_i)  = \prod_{i=1}^L \langle AB_i d^2 A_i \rangle \langle
AB_i d^2 B_i \rangle M_L(AB_i)
\end{equation}
Then, we claim that when the $(AB)_i$ are taken to lie inside
the amplituhedron,
\begin{equation}
M_L(AB_i) > 0
\end{equation}
We will return to exploring this fact at greater in length in
\cite{withandrew}. We stress that this property is {\it not}
manifest term-by-term in the amplituhedron triangulation expansion
of the integrand. Random forms of the local expansion also don't
make this remarkable property manifest term-by-term, but there are
particularly nice forms of local expansion that do make this
manifest. As we will discuss in \cite{withandrew}, we suspect that
this surprising positivity property of the integrand is pointing the
way to a more direct and intrinsic, triangulation-independent
definition for the canonical form $\Omega$ associated with the
amplituhedron.

From a mathematical point of view, it is interesting that the study
of amplitudes leads to stratifications of various collections of
objects in projective space. If we consider a collection of $n$
vectors in $k$ dimensions, together with a cyclic structure on this
data, we are led to the beautiful stratification of the space given
by the positive Grassmannian. Even just with the the four-particle
amplituhedron, we see something new, not needing a cyclic structure
on the objects:  given a collection of $L$ 2-planes in 4 dimensions,
the positivity conditions are fully permutation invariant between
the $L$ lines. Just as with the positive Grassmannian, it is natural to expect the cell structure of
the amplituhedron to be determined in a fundamentally combinatorial
way. The fascinating path-dependence of the forms associated with
the cuts, together with the combinatorics that arise just in the
simple discussion of the multi-collinear limit, are perhaps
indications of an underlying combinatorial structure.

The four-particle amplitude is a truly remarkable object. At the
level of the integrand, at multi-loop order it contains non-trivial
information about all the more complicated multi-particle amplitudes
in the theory. At the level of the final integrated expression, we
have a function that smoothly interpolates between a picture of
``interacting gluons" at weak coupling to ``minimal area surface in
AdS space"\cite{DCI2} at strong coupling. We have explored a reformulation of
this physics in terms of a simple to define, yet rich
and intricate geometry. We hope that this will lead us a more direct
understanding of how the picture of ``gluons"  and ``strings" arise
as different limits of a single object. As a small step in this
direction, it is encouraging to find a natural understanding,
intrinsic to the geometry, of the behavior of the amplitude in the
multi-collinear region, and an associated intrinsic-to-the-geometry
rationale for taking the log of the amplitude. Trying to more
completely determine the IR singular behavior of the integrand of
the amplitude is an ideal laboratory to connect our approach to the
loop integrand with the final integrated expressions, and especially
to ideas related to integrability. Indeed the coefficient of the
log$^2$ infrared divergence of the log of the  amplitude is given by
the cusp anomalous dimension, which was brilliantly determined using
integrability in \cite{BS, Beisert:2006ez,Eden:2006rx}. It is
notable that this approach makes crucial use of a spectral
parameter, something which is absent in our present discussion  of
the amplituhedron. Given the spectral deformation of
on-shell diagrams given in \cite{Ferro:2013dga,Ferro:2012xw}, it is
natural to ask whether a similar deformation can be found directly at the
level of the amplituhedron.

\section*{Acknowledgements}

We thank Jake Bourjaily, Freddy Cachazo, Simon Caron-Huot, Johannes
Henn, Andrew Hodges, Jan Plefka, Dave Skinner and Matthias
Staudacher for stimulating discussions. N.~A.-H. is supported by the
Department of Energy under grant number DE-FG02-91ER40654. J.~T. is
supported in part by the David and Ellen Lee Postdoctoral
Scholarship and by DOE grant DE-FG03-92-ER40701 and also by NSF
grant PHY-0756966.

\end{document}